\documentclass[%
 aip,
 jcp,%
 amsmath,amssymb,
 reprint,%
]{revtex4-1}

\usepackage{graphicx}
\usepackage{dcolumn}
\usepackage{bm}

\begin{document}

\title{Path integral Monte Carlo simulations of dense carbon-hydrogen plasmas}
\author{Shuai Zhang}
\email{shuai.zhang01@berkeley.edu}
\affiliation{Department of Earth and Planetary Science, University of California, Berkeley, California 94720, USA}
\affiliation{Lawrence Livermore National Laboratory, Livermore, California 94550, USA}
\author{Burkhard Militzer}
\email{militzer@berkeley.edu}
\affiliation{Department of Earth and Planetary Science, University of California, Berkeley, California 94720, USA}
\affiliation{Department of Astronomy, University of California, Berkeley, California 94720, USA}
\author{Lorin X. Benedict}
\affiliation{Lawrence Livermore National Laboratory, Livermore, California 94550, USA}
\author{Fran\c{c}ois Soubiran}
\affiliation{Department of Earth and Planetary Science, University of California, Berkeley, California 94720, USA}
\author{Kevin P. Driver}
\affiliation{Department of Earth and Planetary Science, University of California, Berkeley, California 94720, USA}
\affiliation{Lawrence Livermore National Laboratory, Livermore, California 94550, USA}
\author{Philip A. Sterne}
\affiliation{Lawrence Livermore National Laboratory, Livermore, California 94550, USA}
(Accepted for publication in {\it J. Chem. Phys.})
\date{\today}

\begin{abstract}
{Carbon-hydrogen plasmas and hydrocarbon materials are of broad interest to laser shock experimentalists, high energy
density physicists, and astrophysicists.
Accurate equations of state (EOS) of hydrocarbons are valuable for various studies
from inertial confinement fusion (ICF) to planetary science.
By combining path integral Monte Carlo (PIMC) results at high temperatures and density 
functional theory molecular dynamics (DFT-MD) results at lower temperatures, 
we compute the EOS for hydrocarbons from simulations performed at 1473 separate ($\rho,T$)-points distributed over a range of compositions.
These methods accurately treat electronic excitation 
effects with no adjustable parameter nor experimental input.
PIMC is also an accurate simulation method that is capable of treating 
many-body interaction and
nuclear quantum effects at finite temperatures.
These methods therefore provide a benchmark-quality EOS
that surpasses that of semi-empirical and Thomas-Fermi-based 
methods in the warm dense matter regime.
By comparing our first-principles EOS to 
the LEOS 5112 model for CH, we validate the
specific heat assumptions in this model but suggest that
the Gr\"{u}neisen parameter is too large at low temperature.
Based on our first-principles EOS, we predict the principal Hugoniot 
curve of polystyrene to be 2-5$\%$ softer at maximum shock compression than that
predicted by orbital-free DFT and SESAME 7593.
By investigating the atomic structure and chemical bonding of hydrocarbons, 
we show a drastic decrease in the lifetime of chemical bonds
in the pressure interval from 0.4 to 4 megabar.
We find the assumption of linear mixing to be valid for describing the
EOS and the shock Hugoniot curve of hydrocarbons in the regime of partially ionized atomic liquids.
We make predictions of the shock compression of glow-discharge polymers
and investigate the effects of oxygen content and C:H ratio on its Hugoniot curve.
Our full suite of first-principles simulation results may be used to benchmark future theoretical investigations 
pertaining to hydrocarbon EOS, and should be helpful in guiding the design of 
future experiments on hydrocarbons in the gigabar regime.}
\end{abstract}



\maketitle

\section{Introduction}
Accurate equations of state (EOS) of materials under various temperature and 
pressure conditions is of fundamental importance in earth and planetary science, 
astrophysics, and high energy density physics.~\cite{Tang2017} 
These applications require that we understand matter that is partially 
ionized and strongly coupled.
This includes conventional condensed matter ($T \ll T_{\rm Fermi}$), warm dense matter ($T \sim T_{\rm Fermi}$), and weakly coupled plasmas ($T \gg T_{\rm Fermi}$ and $\langle$potential energy$\rangle \ll \langle$kinetic energy$\rangle$). At high temperatures ($>$10$^2$ eV) and 
 near-ambient densities, electrons and nuclei can generally both be treated as ideal gases, due 
to complete ionization of atoms into a perfect plasma state. At slightly lower
temperatures, the Debye-H\"{u}ckel model may be used to treat weak interactions 
within a screening approximation. Materials in condensed forms 
are characterized by both strong coupling and degeneracy effects and thus 
require sophisticated quantum many-body methods for their 
description. However, very good progress can be made with average-atom methods such as average-atom 
Thomas-Fermi theory and average-atom Kohn-Sham Density Functional Theory (DFT), in which isolated ions are embedded within a spherically-symmetric electron liquid. Kohn-Sham-DFT average-atom methods resolve distinct electronic shells of atoms, just as in conventional applications of quantum theory to isolated atoms, but they do not account for directional bonding between atoms and therefore fail at low temperatures. Thomas-Fermi and more general orbital-free (OF) DFT approaches 
do not account for electronic shell effects and are thus unable to properly 
describe partially ionized plasmas; as such, they provide a less than satisfactory description of warm dense matter. 

EOS models (e.g., of the QEOS type~\cite{QEOS}) treating wide ranges of density and temperature and databases that house them (e.g., SESAME~\cite{sesame} and LEOS), make heavy use of average-atom theories for electronic excitations, {\it ad hoc} interpolation formulas which mediate the evolution of the ionic specific heat from low-$T$ to the high-$T$ ideal gas limit, and semi-empirical models which allow for the fitting of experimental results near ambient conditions. The efficacy of these EOS models is questionable precisely in the regimes currently probed in dynamic compression experiments reaching gigabar (Gbar) pressures, in which it is expected that atoms are significantly (though not fully) ionized by both 
temperature and pressure. Clearly, this regime is in need of more sophisticated theoretical treatments which more fully account for detailed electronic structure and many-body effects.


First-principles molecular dynamics (MD) based on DFT
 is widely used for calculating the atomic structure, the EOS, 
and other electronic and ionic properties of materials at 
relatively low temperatures, where the ionization fraction is small.
In DFT-MD, the nuclei are usually treated as classical particles whose motion follows
Newton's equation of motion, whereas the potential field is determined by solving
a single-particle mean-field equation self-consistently using DFT.
This method naturally includes the anharmonic terms of nuclear vibrations 
and is usually a good approximation for systems with heavy elements, and 
is widely used for simulations of earth and planetary materials~\cite{Wilson2010,Wilson2012,Wilson2013,Wahl201525,Zhang2016a,Soubiran2017}.
For light elements, zero-point motion cannot be neglected~\cite{Morales2013a,Morales2013b},
which makes important contributions to the energy that may alter relative phase stability.
At temperatures 
above 100 eV, DFT-MD of the Kohn-Sham variety becomes computationally intractable due to the considerable number of high-lying single-electron states that 
need to be included.
It is also practically limited by the use of pseudopotentials that reduce
computational costs by freezing inner-shell electrons within ionic cores
that tend to overlap between neighboring atoms if the system is
at significant compression,
leading to errors.

Path integral Monte Carlo (PIMC)~\cite{Ceperley1995,Ce96} offers an 
approach to directly solve the many-body Schr\"{o}dinger equation in a 
stochastic way. It typically treats nuclei and electrons as quantum paths that
evolve in imaginary time, and obtains the energy and other properties 
of a system by solving for the thermal density matrix and computing 
thermodynamic averages within the sampled ensembles.
For Fermionic systems, a suitable nodal structure is required to
restrict the sampling space in order to avoid the sign problem
that arises from antisymmetry of the many-body density matrix. 
Accuracy of the method has been demonstrated by early work on
fully-ionized
hydrogen~\cite{PhysRevLett.73.2145,PhysRevE.63.066404} and
helium~\cite{Militzer2009} plasmas. 
In the past five years, developments by extending free-particle 
nodes~\cite{Driver2012} or implementing
localized orbitals~\cite{Militzer2015Silicon} to construct the nodes
have enabled PIMC studies of a series of heavier elements
and compounds~\cite{Driver2012,Benedict2014C,Driver2015Neon,Militzer2015,Driver2015Oxygen,Militzer2015Silicon,Driver2016Nitrogen,PhysRevB.94.094109,Zhang2016b,Driver2017hedp,Driver2017LiF,Zhang2017,Zhang2017b}.
These works have applied PIMC to EOS calculations at temperatures 
ranging from a few hundred million~K to as low as 2.5$\times10^5$ K.
For first- and second-row elements, PIMC and DFT-MD simulations 
produce consistent EOS results at intermediate temperatures.


While computer simulations with classical nuclei provide sufficiently
accurate predictions for a wide range of thermodynamic conditions, at low
temperature, nuclear quantum effects (NQEs) often play an important
role in predicting observed phenomena~\cite{Cazorla2017}. One
typically compares the inter-particle spacing to the thermal de Broglie
wavelength ($\sim1/\sqrt{mk_{\rm B}T}$) that is inversely proportional to
the square root of mass and temperature, and, therefore cold and low-Z
materials are strongly influenced by NQE. Often such effects are
referred to as zero point motion. In solids, zero point effects are
typically studied with lattice dynamics calculations that derive the
spectrum of vibrational eigenmodes within the quasi-harmonic
approximation. Such calculations rely on the second derivative the
energy with respect to the nuclear positions that can be derived
with theoretical methods of various levels of accuracy ranging from classical
force fields~\cite{Rappe1992}, DFT~\cite{Gonze1997,Baroni2001}, and in
principle also with quantum Monte Carlo calculations~\cite{FM99}. It
is difficult, however, to introduce anharmonic effects accurately into
the lattice dynamics approach~\cite{Zhou2014}. Anharmonic effects are
important for temperatures above the Debye temperature, in particular
near melting, but also at very low temperature in materials that are
rich in hydrogen or helium. Because of their small mass, the nuclear
wavefunctions often spread into the anharmonic regions of the confining
potential even in the ground state so that anharmonic effects can no 
longer be neglected.

Path integral methods~\cite{Fe72} can accurately incorporate NQE and anharmonic
effects at all temperatures and can describe solids as well as 
liquids~\cite{Jo96,Militzer20062136}. An efficient approach, that has been devised
to pursue NQE problems, is to combine the path integral method for
nuclei with other electronic structure methods, such as DFT or quantum
Monte Carlo, to efficiently describe the forces between the nuclei.
Path integral molecular dynamics~\cite{KiDongOh1998} and coupled
electron ion Monte Carlo~\cite{Pierleoni2006} are two common
techniques that employ this combined approach~\cite{WDMBook2014}.  A
few examples include 
 transport properties~\cite{Kang2014} and
phase transitions in
solid~\cite{McMahon2012,Morales2013b} and
liquid~\cite{Morales2010,Morales2013a,Pierleoni2016} hydrogen,
helium~\cite{Pollock1984,Militzer2009}, hydrogen-bonding in
water~\cite{Chen2003,Morrone2008,Li2011,Ceriotti2016} 
and on metal surfaces~\cite{Li2010}, 
solid water-ice
phases~\cite{Benoit1998,Pamuk2012}, phonon dispersion in
diamond~\cite{Ramirez2006} and energy barriers for small
molecules~\cite{Weht1998,Hauser2017}. 

In this work, we employ the path integral methods to study partially
and highly excited electrons. Even though the temperatures under
consideration are quite high ($\sim$100 eV), fermionic effects are crucial
to characterize the electronic states accurately. The occupation of
bound electronic states affects the motion of the nuclei through Pauli
exclusion. At high density, the motion of the nuclei deforms the shape
of the electronic orbitals. Thus, a fully self-consistent approach is
needed for materials in the WDM regime.


We apply PIMC simulations with free-particle nodes and 
DFT-MD calculations to 
study carbon-hydrogen compounds. Hydrocarbons are currently in use as ablator 
materials in dynamic compression and inertial confinement fusion (ICF) experiments~\cite{Betti2016,Meezan2016,Goncharov2016,Guillot2005,Wallerstein1997}.
Materials such as polystyrene and glow-discharge polymer (GDP)
have been 
a strong point of interest and extensively studied by both theorists and experimentalists~\cite{Hauver1964,Dudoladov1969,Lamberson1972,Marsh1980,Nellis1984,Kodama1991,Bushman1996,Koenig1998,Cauble1997,Cauble1998,Koenig2003,Ozaki2005,Hu2008,Ozaki2009,Barrios2010,Shu2010,Shang2013,Shu2015,Barrios2012,Huser2013,Huser2015,Moore2016,Hamel2012}.
Recently, laser shock experiments at the National Ignition
Facility (NIF)~\cite{Kraus2016,Swift2011,Doppner2014,Kritcher2014,Kritcher2016,Nilsen2016} 
and the OMEGA Laser facility~\cite{Nora2015} have extended 
the ablation pressure in CH to 
the Gbar range~\cite{CHGbarExpts}.
Calculations employing DFT-MD or OF-DFT have been performed
to study the EOS of hydrocarbon materials~\cite{Mattsson2010,Wang2011,Lambert2012,Hu2014,Hu2015,Hu2016,Huser2015,Colin2016}.
These theoretical studies predicted shock Hugoniot curves that agree well 
with experiments at low temperatures. At high temperatures, Kohn-Sham DFT
is unfeasible while OF-DFT works efficiently but its predictions are yet to be
tested by other theories~\cite{PhysRevB.94.094109} and experiments.
We expect the PIMC and DFT-MD simulations of this study to 
produce predictions for the EOS of hydrocarbons 
which are sufficiently accurate to be used as benchmarks 
for both the construction of wide-range EOS models, 
and the further investigation of hydrocarbon EOS with
 less computationally expensive simulation methods. 
Ultimately, our predictions may prove useful for the design and 
interpretation of dynamic compression and ICF experiments in the future.

The paper is organized as follows: Section~\ref{method} introduces the details of
our simulation methods.  Sec.~\ref{results} presents our EOS results, 
the shock Hugoniot curves, and comparisons with other theories, models, and experiments. 
Sec.~\ref{discuss} discusses the structural evolution of hydrocarbons and the 
shock compression of GDP and related materials. We conclude in Sec.~\ref{conclusion}.

\section{Simulation methods}\label{method}
We use the {\footnotesize CUPID} code~\cite{militzerphd} for our PIMC
simulations within the fixed-node approximation~\cite{Ceperley1991}.
We treat the nuclei as quantum particles, even though the kinetic
energy is much larger than the zero-point contribution to the total
energy of the system at the high temperatures ($T\ge10^6$ K = 87 eV)
considered here.   Electrons are treated as fermions. Their quantum
paths are periodic in the imaginary time interval, $0\le t\le\beta=1/k_\text{B}T$
($k_\text{B}$ is the Boltzmann constant), but the paths of electrons
with the same spin may be permuted as long as they do not
violate the nodal restriction~\cite{Ceperley1991}. Following our
previous work on
hydrogen~\cite{PhysRevLett.73.2145,PhysRevLett.76.1240,PhysRevE.63.066404,PhysRevLett.87.275502,PhysRevLett.104.235003,PhysRevLett.85.1890,PhysRevB.84.224109,CTPP:CTPP2150390137,Militzer20062136}, helium~\cite{PhysRevLett.97.175501,Mi05},
and carbon~\cite{Driver2012,Benedict2014C}, we use free-particle nodes to constrain
the sampling space by restricting the paths to positive regions of the
density matrix of ideal fermions. Coulomb interactions between all
pairs of particles are introduced via pair density
matrices~\cite{Na95,pdm}. The pair density matrices are evaluated at an
imaginary time interval of 1/1024 Hartree$^{-1}$ (Ha$^{-1}$) while the
nodal restriction is enforced in steps of 1/8192 Ha$^{-1}$.

For DFT-MD simulations, we use the Vienna \textit{Ab initio} Simulation Package
({\footnotesize VASP}) \cite{kresse96b}. 
We choose the hardest available projected augmented wave (PAW)
pseudopotentials \cite{Blochl1994} with core radii
of 1.1 and 0.8 Bohr for C and H, respectively. All electrons are treated as valence electrons.
Exchange-correlation effects are treated within the 
local density approximation (LDA)~\cite{Ceperley1980,Perdew81}.
We choose a large plane-wave basis cutoff of 2000 eV,
the $\Gamma$ point to sample the Brillouin zone,
and an MD time step of 0.05-0.2 fs, depending on the temperature.
MD trajectories are generated in an $NVT$ ensemble,
and typically consist 2000-10000 steps to make sure the system 
is in equilibrium and the energies and pressures are converged.
The temperature is controlled with a Nos\'{e} thermostat~\cite{Nose1984}.
All {\footnotesize VASP} energies are shifted by -37.4243~Ha/C and 
-0.445893~Ha/H, to put DFT-MD energies on the same scale as 
our all-electron PIMC energies.
 These values are determined by performing
all-electron single-atom calculations using the {\footnotesize OPIUM} code~\cite{opium}.

\begin{figure}
\centering\includegraphics[width=0.5\textwidth]{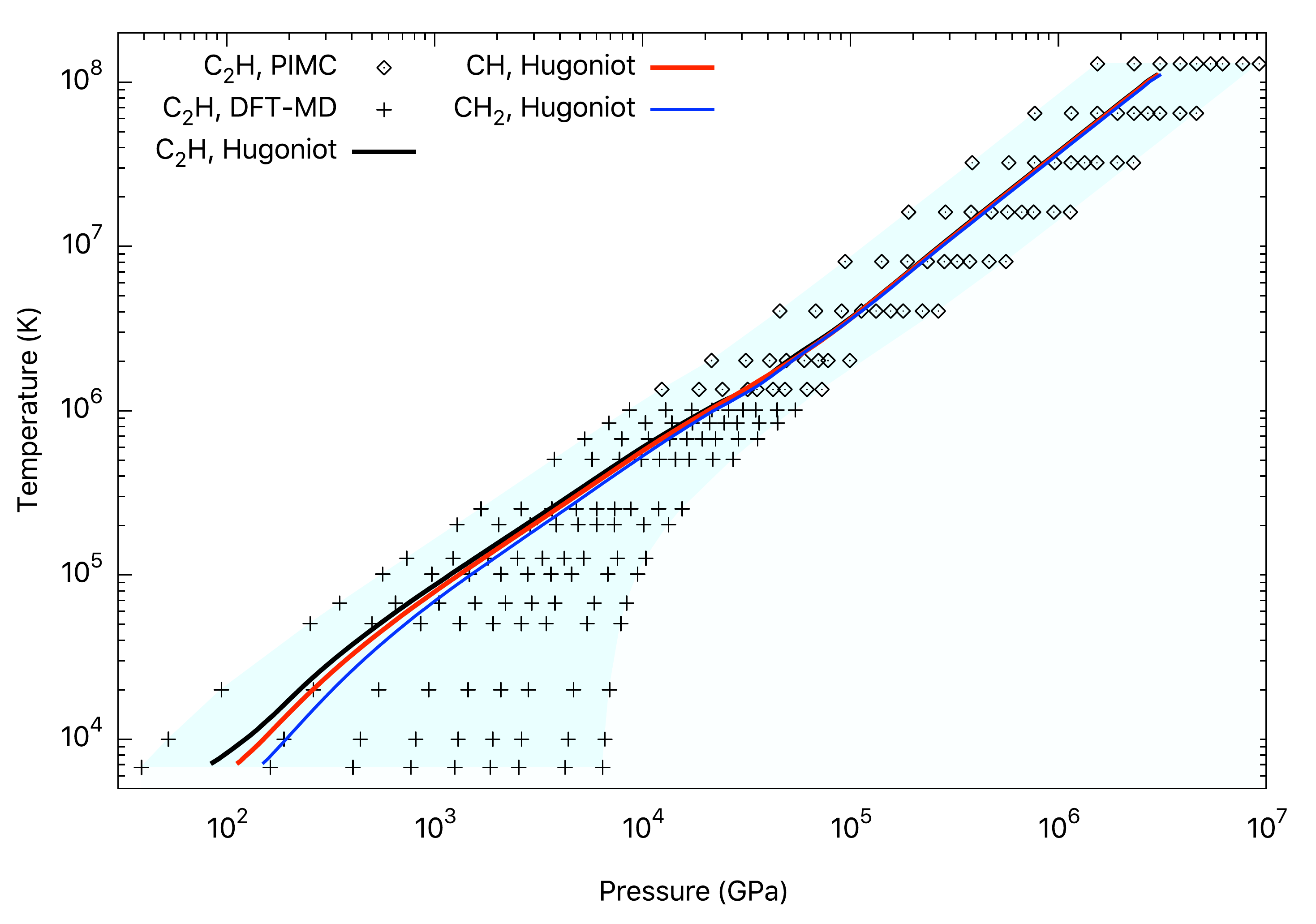}
\caption{\label{cmhneos}  Pressure-temperature conditions 
considered in our PIMC and DFT-MD simulations of hydrocarbons. For clarity, 
only the EOS data for C$_2$H and the Hugoniot curves of C$_2$H, CH, and CH$_2$ 
are shown.
The Hugoniot curves are obtained by setting the initial denstity
to 1.12, 1.05, and 0.946~g/cm$^3$ for C$_2$H, CH, and CH$_2$, respectively.
The shaded region denotes the approximate area that the EOS of all hydrocarbons 
studied in this work fall in.
}
\end{figure}

We study six different compositions by simulating C$_{20}$H$_{10}$, C$_{18}$H$_{18}$,
C$_{16}$H$_{24}$, C$_{14}$H$_{28}$, C$_{12}$H$_{36}$, and C$_{10}$H$_{40}$ in a cubic cell at temperatures
between 10$^6$-1.3$\times10^8$ K using PIMC and 
6.7$\times10^3$-10$^6$ K using DFT-MD methods.
For C:H=1:1, we use larger cells with four times as many atoms at the lower temperatures 
of 6.7$\times10^3$-2.5$\times10^5$ K in order to minimize finite-size effects,
which are expected to be larger at low temperatures~\cite{Driver2012}.
In order to maximize the computational efficiency of the large-cell simulations,
we freeze the 1s$^2$ electrons of carbon in the pseudopotential core
without losing accuracy, because the temperatures are much lower than
the ionization energy (392~eV for 1s$^2$ of C)~\cite{nistEion}.
We consider a grid of nine isochores for each hydrocarbon system, 
chosen so that the pressure ranges for each composition are similar; 
this results in densities for CH between 
(2 - 12)$\times$$\rho_\text{ambient}$
(see Fig.~\ref{cmhneos}).
These conditions are both relevant to dynamic compression experiments, and well within the range in which Kohn-Sham DFT-MD simulations with pseudopotentials are feasible. 

We also investigate the validity of the linear mixing approximation for
estimating the EOS of various hydrogen-carbon mixtures. 
In this method, the energy and the density of the mixture are obtained with the isobaric, isothermal  additive volume assumption via $V_{\rm mix}(P,T)= f_{\rm C}V_{\rm C}(P,T) + f_{\rm H}V_{\rm H}(P,T)$ and $E_{\rm mix}(P,T)= f_{\rm C}E_{\rm C}(P,T) + f_{\rm H}E_{\rm H}(P,T)$, where $f_{\rm C}= n_{\rm C}/(n_{\rm C} + n_{\rm H})$ and $f_{\rm H}= n_{\rm H}/(n_{\rm C} + n_{\rm H})$ are the 
mixing ratios, $V_{\rm mix}$, $V_{\rm C}$, $V_{\rm H}$ are volumes per atom for the mixture, pure carbon,
and pure hydrogen respectively (and likewise for the internal energies, $E$). 
$E(T,P)$ and $V(T,P)$ of the pure species are constructed using bi-variable spline fitting 
over the $\rho$-$T$ space spanned by the EOS of pure hydrogen and pure carbon. The shock Hugoniot curves derived with simulations of the fully interacting system can be compared
to that obtained with the linear mixing approximation.

\section{Results}\label{results}

\subsection{Equation of state}\label{seceos}

\begin{figure}
\centering\includegraphics[width=0.5\textwidth]{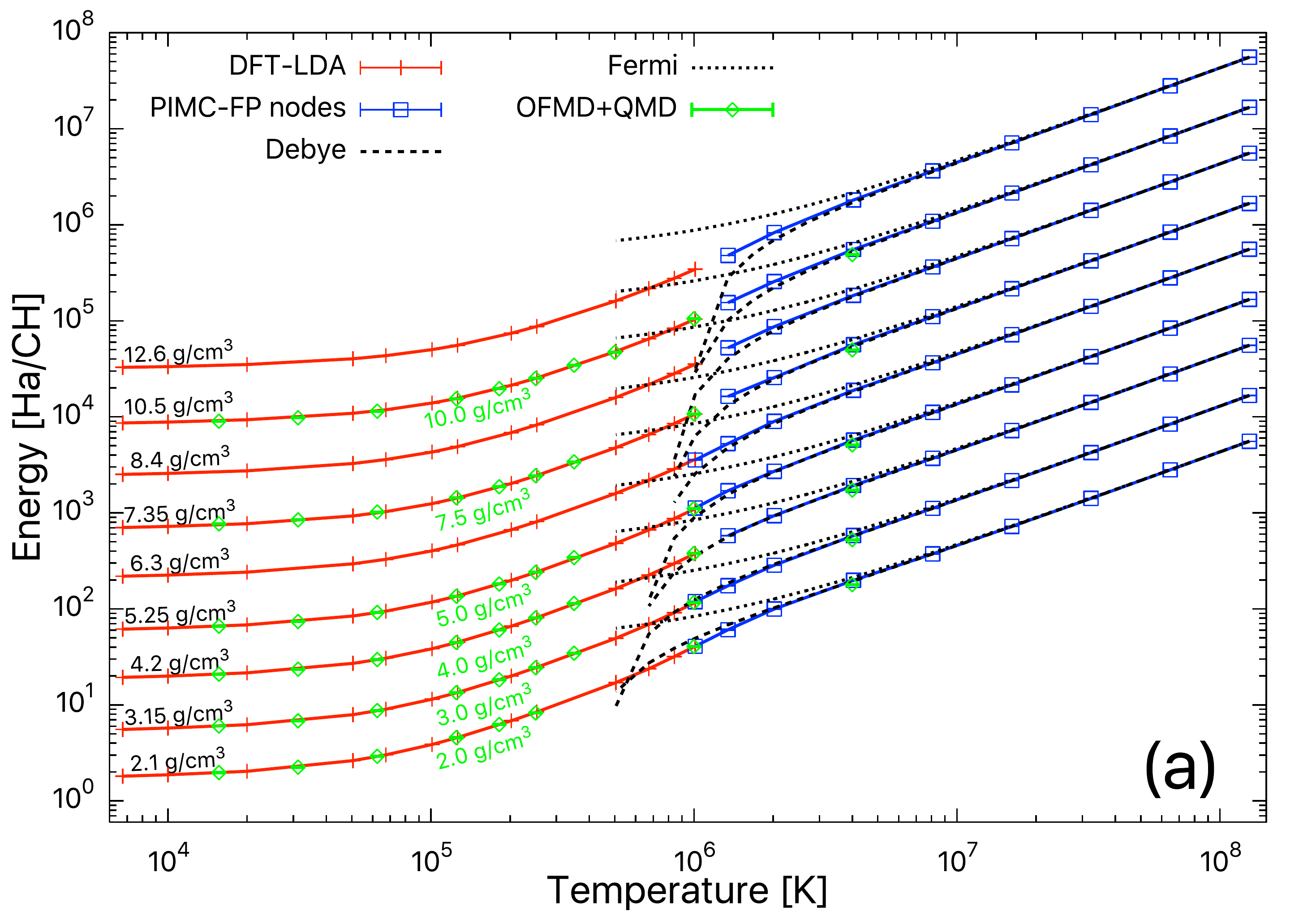}
\centering\includegraphics[width=0.5\textwidth]{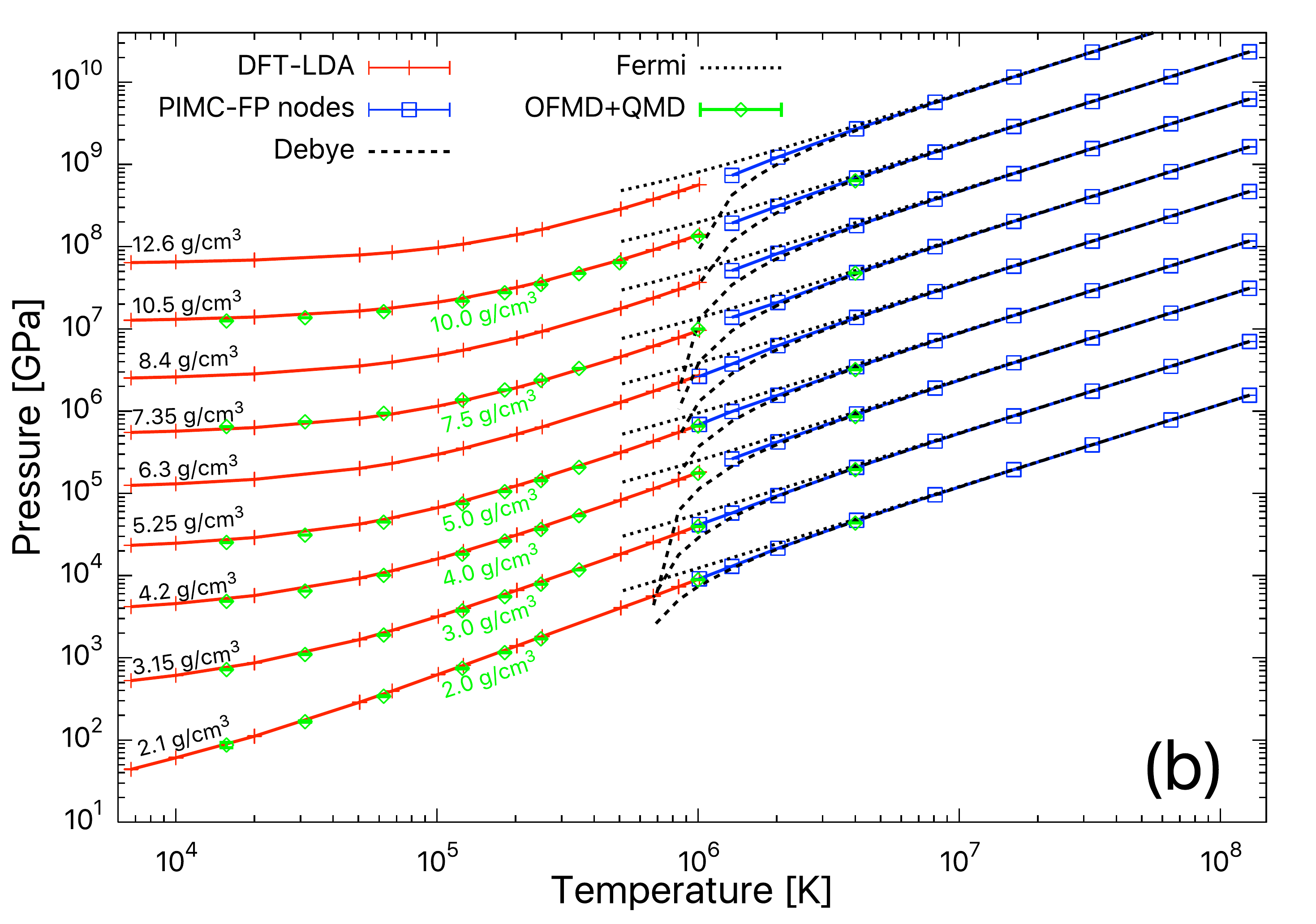}
\caption{\label{chvsofmd} Internal energies and pressures of CH
along isochores. Corresponding
results of the Fermi-gas and the Debye-H\"{u}ckel theories are plotted for
comparison.
PIMC predicts consistent results with DFT-MD
at 10$^6$ K PIMC and agrees with ideal Fermi gas and Debye-H\"{u}ckel
models at above 8$\times$10$^6$ K.
As temperature decreases, the ideal Fermi gas model
significantly overestimates the energy and the pressure
because of the neglect of interactions.
Debye-H\"{u}ckel model improves over the Fermi-gas model
for temperatures down to 3$\times$10$^6$ K
but leads to low pressures and energies at lower temperatures, at which
electron-nucleus coupling gets stronger and the screening approximation breaks down.
DFT-MD and OF-DFT results from Ref.~\onlinecite{Hu2015}
are shown with green diamonds for comparison.
Error bars of the data are much smaller than the size of the symbols.
Different isochores have been shifted apart for clarity. }
\end{figure}

Figure \ref{chvsofmd} shows the calculated EOS for C$_2$H
along nine isochores. A complete list of EOS data for all
hydrocarbons (C$_2$H, CH, C$_2$H$_3$, CH$_2$, CH$_3$, CH$_4$) 
in this study are in the supplementary material.
The internal energies and pressures from PIMC calculations 
 agree with predictions of the Debye-H\"{u}ckel model at temperatures 
above 4$\times10^6$ K and with the Fermi electron gas theory (wherein both ions and electrons are treated as uniform-density free Fermi gases)
above 8$\times10^6$ K,
which is higher than the 1s$^1$ ionization energy 
(489.99 eV or 5.7$\times10^6$ K)~\cite{nistEion} of carbon.
PIMC results show excellent agreement with DFT-MD at 10$^6$ K, 
with differences typically
less than 1~Ha/carbon in internal energy and 3\% in pressure. 
We have therefore constructed a consistent first-principles EOS
table for warm dense hydrocarbons, over a wide density-temperature
range of 1.4-13.5 g/cm$^3$ and $6.7\times10^3$-$1.3\times10^8$ K
and C:H=2:1-1:4. 
The good consistency of PIMC with DFT-MD and the other high-temperature
theories indicates it is reliable to use the 
free-particle nodes in PIMC for
temperatures as low as $10^6$ K, and 
PAW pseudopotentials and
zero-temperature exchange-correlation functionals in DFT-MD up to $10^6$ K.

In comparison with a recent first-principles study~\cite{Hu2015} that
employs DFT-MD with the Perdew-Burke-Ernzerhof (PBE)~\cite{Perdew96} exchange-correlation 
functional at temperatures below the Fermi temperature $T_{\rm Fermi}$, 
the $P$-$T$ and $E$-$T$ curves coincide with LDA curves in this work.
This indicates the EOS does not significantly depend on the form of the
exchange-correlation functional in the temperature interval under consideration. 
At $10^6$ K and higher temperatures, 
the energy of Ref.~\onlinecite{Hu2015} is different from PIMC 
predictions of this work while the pressure differences are small.
This is associated with underestimation of the compression maximum
by OF-DFT that is used in Ref.~\onlinecite{Hu2015}.
More details on this will be discussed in Sec.~\ref{secshock}.

\subsection{Comparisons with the LEOS-5112 model for CH}
An important aim of this work is to produce benchmarking EOS predictions that will 
act as constraints in
 the future construction of EOS models for hydrocarbons which span wide ranges of density and temperature, well beyond that where experimental data is available. It is therefore interesting to compare the details of {\it existing} EOS models for, e.g., CH to the first-principles predictions in this work. The most recent such model for polystyrene (CH) constructed at the Lawrence Livermore National Laboratory is LEOS-5112, 
closely related to LEOS-5400 \cite{Sterne2016}, currently used as the EOS model of choice for GDP in inertial confinement fusion simulations where that material is employed as an ablator. EOS models such as these assume that the free energy is decomposable into separate ionic and electronic excitation terms. While somewhat justified due to the large ion/electron mass ratio, it is important whenever possible to compare to EOS predictions from {\it ab initio} methods (such as PIMC) that do {\it not} make this assumption. 

The LEOS-5400 EOS model~\cite{Sterne2016} was originally constructed to represent the non-stoichiometric carbon-hydrogen-oxygen composition of GDP, and to reproduce both Hugoniot and off-Hugoniot measurements (specifically, shock-and-release from an interface with deuterium)~\cite{Hamel2012}.  
More recently, LEOS-5112 was developed to facilitate comparison with experiments and simulations on the stoichiometric material, CH. This CH model closely follows the parameters used to make GDP LEOS-5400.
 
The cold curve, $E(V,T=0)$, was based on a constant-pressure mix of the corresponding Thomas-Fermi cold curves for pure C and pure H, with a bonding correction~\cite{QEOS} to set the density at 1.049 g/cm$^{3}$ and the bulk modulus at 0.9 GPa at a temperature of 20 K. This relatively high bulk modulus was softened by using a break point~\cite{Young1995} that changes the cold curve energy by $\delta E= Au^{3}/(B+u^{3})$ for $u \ge 1$ where $u = \rho/\rho_0, A=$-8 kJ/g and $B=3.0$.  A corresponding change was also made to the cold curve pressure. The relatively high bulk modulus value, together with the break point, reproduces both the initial equilibrium conditions of the material (at cryogenic conditions), and the higher-temperature behavior above 50 GPa, which is higher than the graphite-diamond transition pressure  along the Hugoniot ($\sim$15-25 GPa).

\begin{figure}
\centering\includegraphics[width=0.5\textwidth]{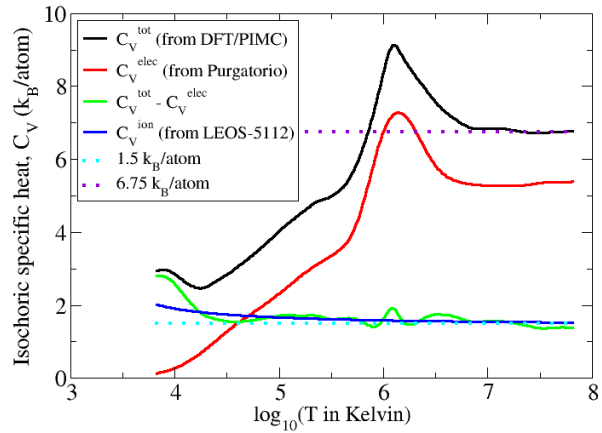}
\caption{\label{lorincv} $C_{V}$ for CH at a density of $\rho= 3.15$ g/cm$^{3}$. The black curve is $C_{V}$ as extracted from a spline fit of our discrete $E(T)$ predictions from first-principles simulations at this density. See the text for descriptions of the other curves.}
\end{figure}
 
\begin{figure}
\centering\includegraphics[width=0.5\textwidth]{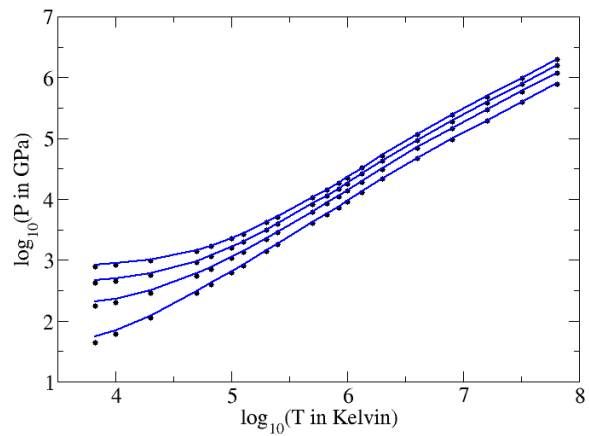}
\caption{\label{lorinp} Total pressure isochores for CH at densities $\rho=$ 2.1, 3.15, 4.2, and 5.25 g/cm$^{3}$ (in the sequence of bottom to top). The black symbols are the results of our 
DFT-MD ($T\le10^{6}$ K) and PIMC ($T > 10^{6}$ K) simulations. The blue curves are the corresponding pressure isochores from LEOS-5112.}
\end{figure}

The terms in the free energy accounting for ionic excitations (ion-thermal) were modeled using both a Debye-Gr\"{u}neisen model and a dissociation model.  The Debye model used a Debye temperature of 650~K at equilibrium density and an ion-thermal Gr\"{u}neisen $\gamma$ at this point of 0.99.  This value of $\gamma$ was originally chosen to give the experimentally observed thermal expansion of GDP between 20 K and room temperature. The Gr\"{u}neisen parameter was kept constant 
from $\rho_0$
up to a density of 2.7 g/cm$^{3}$, at which point it was gradually decreased to 0.81 at 10 g/cm$^{3}$ and 0.5 at very high density. 
For densities below 1.049 g/cm$^{3}$, the Gr\"{u}eneisen gamma reduces gradually to the ideal-gas value of 2/3.
The variation in the Debye temperature is computed from this $\gamma(\rho)$ function~\cite{QEOS}.  The dissociation model adds an additional contribution to the free energy which models the dissociation of a dimer~\cite{Young1995}.  The main purpose of this term is to model chemical dissociation in a simplified way, as if it were due to diatomic molecular dissociation.  In the GDP model, this extra flexibility allowed simultaneous fits to both Hugoniot data and off-Hugoniot release data.  This same model was retained in LEOS-5112 without modification.  The model includes a dimer dissociation energy of 0.7 eV and a nominal rotational temperature of 20 K.  

The liquid contribution to the ion-thermal free energy in LEOS-5112 (and LEOS-5400) is given by a Cowan model \cite{QEOS} with an exponent of 1/3, 
\begin{equation}
C_{V}^{\rm ion}= \frac{3k_{B}}{2} + \frac{3k_{B}}{2}\left[\frac{T_{m}(\rho)}{T}\right]^{1/3}
\label{Cv}
\end{equation}
This ensures that the ionic contribution to the specific heat decays from $3k_{B}/$ion to the ideal gas value of 1.5~$k_{B}/$ion as $T\rightarrow \infty$.  $T_{m}(\rho)$ is taken to be the melt temperature, determined from the Lindemann relation \cite{QEOS}. 
Since CH dissociates before melting, it is necessary to assume a value 
for the melt temperature at ambient pressure. 
The value used here, $T_{m}$=513.15 K, is the same as that used in LEOS-5400 for GDP.
 
The electronic excitation contribution to the free energy of LEOS-5112 comes from the Purgatorio atom-in-jellium calculations~\cite{Purgatorio2006} for carbon and hydrogen.  The Purgatorio cold curve is replaced by a Thomas-Fermi cold curve, and some data adjustment is done in the low temperature region around equilibrium density to guarantee monotonicity in the pressure.  These tables are then mixed using a constant-pressure, constant-temperature additive volume mix procedure.  The resulting cold curve is then subtracted to yield the electron-thermal contribution used in the EOS.  This includes the effects of shell structure, as well as relativistic effects at very high temperatures.  

Figure~\ref{lorincv} shows the specific heat at constant volume, $C_{V}$, for CH at a density of 3.15~g/cm$^{3}$. The black curve is the result of calculating $(\partial E/\partial T)_{V}$ directly from the DFT-MD (for $T \le 10^{6}$ K) and PIMC (for $T > 10^{6}$ K) internal energies, by fitting a cubic spline to our 20 $E(T)$ points at this density and differentiating the smooth spline function. At the highest $T$, this asymptotes to 6.75 $k_{B}$/atom, which is the required value from equipartition, assuming complete ionization. There is a notable peak in this curve just above $10^6$ K. The red curve shows $C_{V}^{\rm electron}$ from LEOS-5112, obtained as described above from the Kohn-Sham DFT average-atom Purgatorio model~\cite{Purgatorio2006}. Though the peak is in a slightly different position than that seen in the black curve, this suggests that (a) the peak in $C_{V}$ at $T= 10^{6}$ K arises from electronic ionization, and (b) this feature is modeled well by the comparatively simple Pugatorio treatment (which neglects, e.g., directional bonding). 
The temperature interval of the peak and related information presented in
Refs.~\onlinecite{Zhang2017b,PhysRevB.94.094109} suggest that its appearance results primarily from the ionization of the 1s electron shell of carbon.
The green curve shows the black curve minus the red curve, which is an estimation of the ionic contribution to $C_{V}$, assuming the perfect additivity of electronic and ionic components together with the use of Purgatorio for the electronic piece. Note that it approaches the required value of 1.5~$k_{B}$/atom at high-$T$ while rising slowly as $T$ decreases, ultimately rising rapidly to a value of $\sim 3 k_{B}$/atom at the lowest $T$ (a value of $\sim$2.9~$k_{B}$/atom was predicted for CH$_{1.36}$ in Ref.~\onlinecite{Hamel2012}). This rapid rise is reminiscent of behavior seen in DFT-MD calculations of pure carbon (see Fig.~9 of Ref.~\onlinecite{Benedict2014C}) where $C_{V}$ was shown to drop quickly from $3k_{B}$/atom with increasing $T$ over the range 2$\times10^4$-8$\times10^4$ K. Above a few times $10^{4}$~K, the general behavior of the green curve is tracked rather well by the blue curve, which shows $C_{V}^{\rm ion}$ from LEOS-5112 at this density, given by the form in Eq.~\ref{Cv}. With the exception of the small discrepancies at the lowest $T$ of Fig.~\ref{lorincv}, the general agreement shown here provides validation for the manner in which the specific heat is modeled in LEOS-5112, indicating that at least the $T$-dependent part of $E$ is treated reasonably well. This is noteworthy, for the work on pure carbon~\cite{Benedict2014C} demonstrated that the model of Eq.~\ref{Cv} was suspect even for $T$ as high as $10^{6}$ K.
The behavior shown in Fig.~\ref{lorincv} is reproduced in our other energy isochores as well.

\begin{figure}
\centering\includegraphics[width=0.5\textwidth]{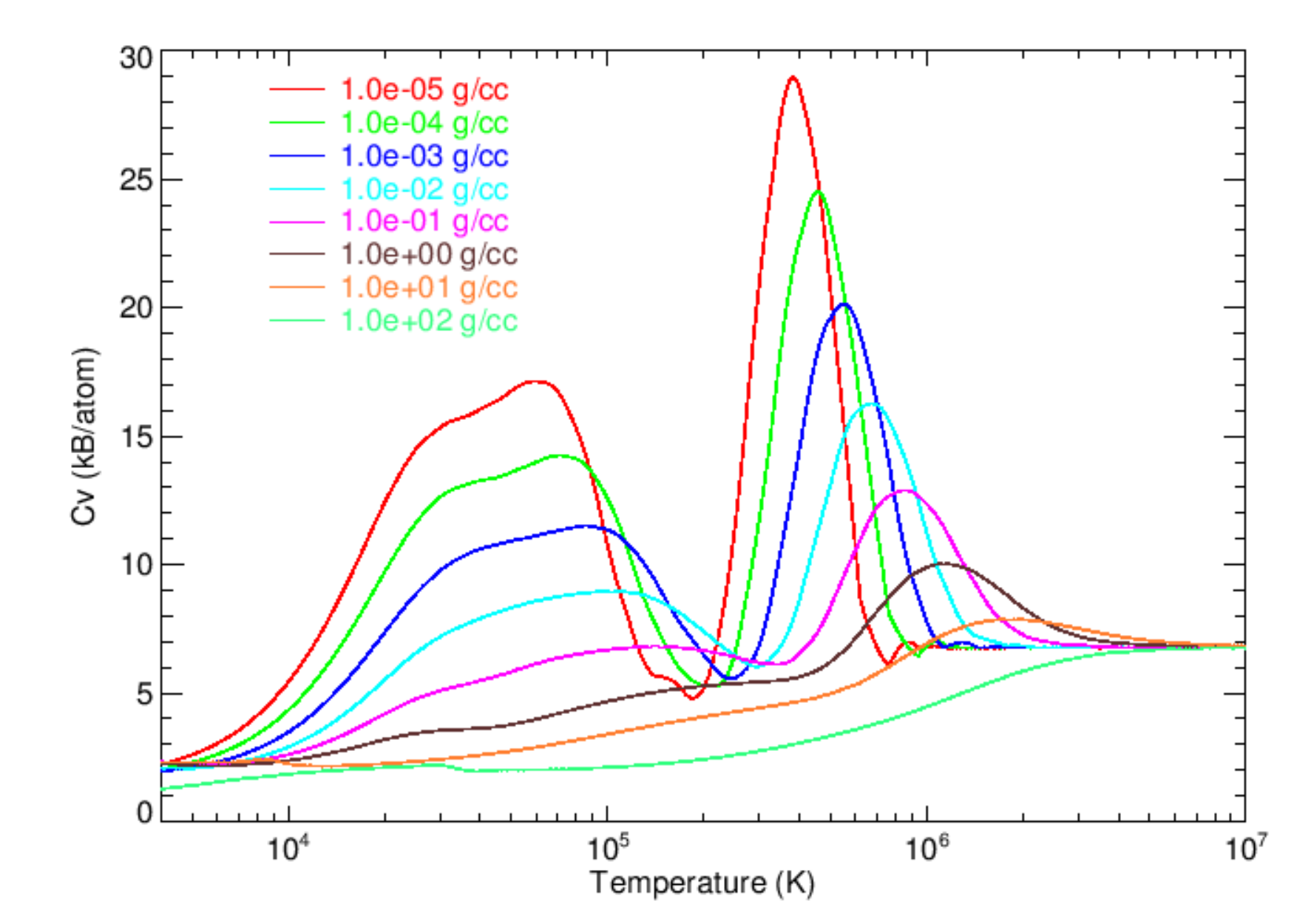}
\caption{\label{cvtrho} $C_V(T)$ of CH from LEOS 5112 at $10^{-5}, 10^{-4}, 10^{-3}, 10^{-2}$, 0.1, 1,  10, and 100 g/cm$^{3}$.}
\end{figure}

We further study the density dependence of $C_{V}(T)$ using LEOS 5112 
in order to better understand the ionization process of CH. 
In Fig.~\ref{cvtrho}, we show a comparison of the
$C_{V}$ profiles over a much wider density range
from $10^{-5}$-100~g/cm$^{3}$ than we can access with existing simulation methods.
For densities $\lesssim$0.1~g/cm$^{3}$, we find two peaks:
a narrow one above $3\times10^5$~K corresponding to the ionization of the K shell (1s state),
and a wide one at 10$^5$~K due to the L shell of the carbon atoms. 
As the density increases, both peaks widen and shift to higher temperatures.
These changes are due to the complex interplay of three effects:
variations in the orbital shapes and an increase in size
relative to the interatomic distances,
a depletion of the 1s Fermi occupancy due to the temperature increase,
and changes in the 1s binding energy due to variations in the screening effects.
The K-shell peak in the heat capacity is predicted to disappear at $\sim$100~g/cm$^{3}$, 
even though the 1s orbital is still bound at this density, and is not fully pressure ionized 
until a density of over 500~g/cm$^{3}$. The suppression of the peak is a  result of changes 
in the 1s binding energy as the 1s state is thermally depopulated. As the Fermi function 
reduces the occupancy of the orbital with increasing temperature, the binding energy of 
the 1s state increases due to a reduction in screening.  This, in turn, raises the temperature 
required to depopulate the state, which results in a broadening of the heat capacity peak, 
leading to its apparent disappearance for the 100~g/cm$^{3}$ case. This reduction-in-screening 
effect is present at all densities, but it is only at high enough densities since the heat capacity 
peak is shifted to sufficiently high temperatures so that it becomes undetectable when broadened.
For densities less than 0.01~g/cm$^3$,
we also identified a shoulder in the L-shell peak at approximately 3$\times10^4$ K, 
which we attribute to ionization effects of the 
hydrogen atoms because this shoulder is absent from
the high-density carbon Purgatorio-based table.
 
While the temperature dependence of the internal energy of LEOS-5112 is therefore in quite good accord with that of our first-principles results throughout a wide range of density and temperature, the pressure of this model shows more significant discrepancies. Figure~\ref{lorinp} shows isochores of $P$ for four densities, $\rho=$ 2.1, 3.15, 4.2, and 5.25 g/cm$^{3}$. 
At the lower-$T$, the pressures of the model are systematically too high; near-perfect agreement is only seen at much higher temperature, approaching the ideal gas limit. This suggests that the Gr\"{u}neisen $\gamma$ is too large in LEOS-5112, since this quantity does not enter the $C_{V}$ discussed above. Still another possibility is that the cold curve of LEOS-5112 can benefit from slight modification, though our current inclination is to reinvestigate the combined Debye-Gr\"{u}neisen and dissociation models pertaining to the ionic excitation contribution. Work to improve the EOS models of hydrocarbons such as CH using these first-principles simulations is ongoing. In the following Sec.~\ref{secshock}, LEOS-5112 and LEOS-5400 are discussed again, in the context of predicted Hugoniot curves. 


\subsection{Shock compression}\label{secshock}

The locus of final states characterized by ($E,P,V$) accessible via a planar one-dimensional shock satisfy the Rankine-Hugoniot energy equation~\cite{Meyers1994book}
\begin{equation}\label{eqhug}
\mathcal{H}(T,\rho)=(E-E_0) + \frac{1}{2} (P+P_0)(V-V_0) = 0,
\end{equation}
where ($E_0,P_0,V_0$) are the variables characterizing the initial (pre-shocked) state.
This allows for the determination of the $P$-$V$-$T$ Hugoniot curve 
with bi-variable spline fitting of the EOS data ($E$ and $P$ on a grid of $T$ and $V$) 
in Sec.~\ref{seceos}.

\begin{figure}
\centering\includegraphics[width=0.45\textwidth]{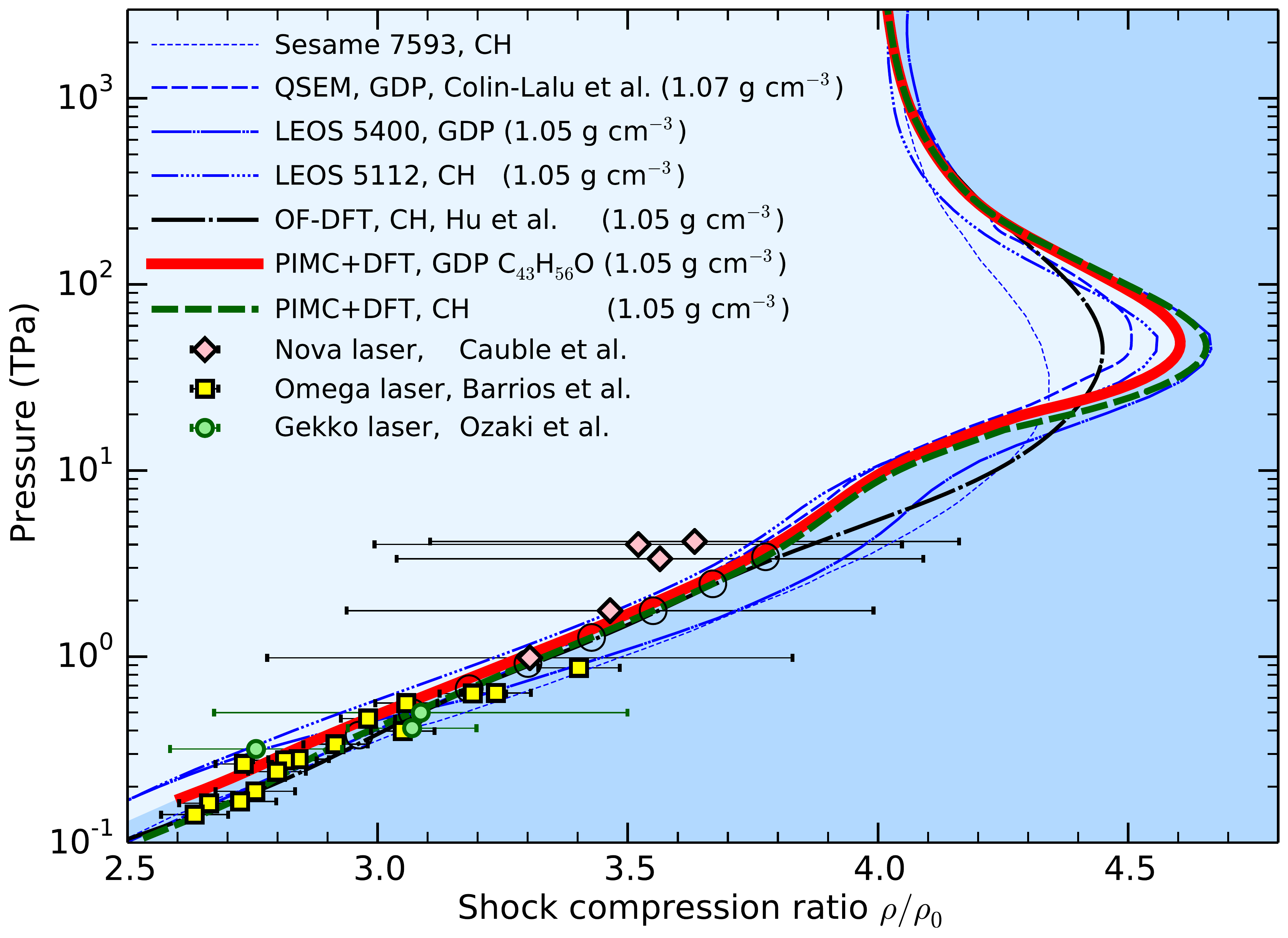}
\caption{\label{fig_huggdp1} Comparison of the shock Hugoniot curve of CH and GDP from first-principles
calculations (PIMC and DFT-MD simulations or linear-mixing approximation in this work, OF-DFT and DFT-MD simulations in Ref.~\onlinecite{Hu2015}), EOS models (SESAME 7593~\cite{sesame}, QSEM~\cite{Colin2016}, LEOS 5400~\cite{Sterne2016}, LEOS 5112), and experiments~\cite{Cauble1997,Cauble1998,Ozaki2009,Barrios2010}.
Black circles are conditions chosen for examining the linear
mixing approximation (see Sec.~\ref{struccharac}).}
\end{figure}

We thus obtain the principal Hugoniot curves of hydrocarbons~\cite{Zhang2017b} 
and represent that of CH (assuming $\rho_{0}= 1.05$ g/cm$^{3}$ and $T_{0}= 300$ K) in a pressure-density plot (Fig.~\ref{fig_huggdp1}) together with experimental 
measurements at low pressures~\cite{Cauble1997,Cauble1998,Ozaki2009,Barrios2010}, OF-DFT simulations~\cite{Hu2015}, and
the Purgatorio-based LEOS 5112 (described above), QSEM~\cite{Colin2016},
and the SESAME 7593~\cite{sesame} models.
The DFT-MD predictions of the Hugoniot curve in this work
and that of Ref.~\onlinecite{Hu2015}
agree well with experimental measurements, whereas SESAME 7593 deviates 
from the experimental data at 1-4 TPa.
At 0.5 Gbar, we predict CH to reach a maximum compression ($\rho/\rho_{0}$) of 4.7, which is
similar to that predicted by the LEOS 5400 model for GDP, and is higher than that of 
QSEM~\cite{Colin2016}, LEOS 5112, and OF-DFT~\cite{Hu2015} by 2-5\%; 
the SESAME 7593 model predicts the maximum compression to be smaller by 7.3\%
and the corresponding pressure about 17 megabar (Mbar) lower than our first-principles simulations.
The compression maximum is originated from the 1s shell ionization of carbon.
Since such temperature and pressure conditions correspond to the region
at which PIMC works well and complexities such as electronic quantum effects, 
electronic correlation, and partial ionization are all essentially
included in the quantum many-body framework,
we expect the predictions in this work to be more reliable than
those of semi-empirical models
and OF-DFT.
We note that the shock Hugoniot curve of CH obtained in this work
is in remarkable agreement with that using a recent extended DFT method~\cite{ZhangExtendedDFT2016,201703communication}.
We look forward to accurate experiments in the Gbar regime which test these predictions.

\begin{figure}
\centering\includegraphics[width=0.35\textwidth]{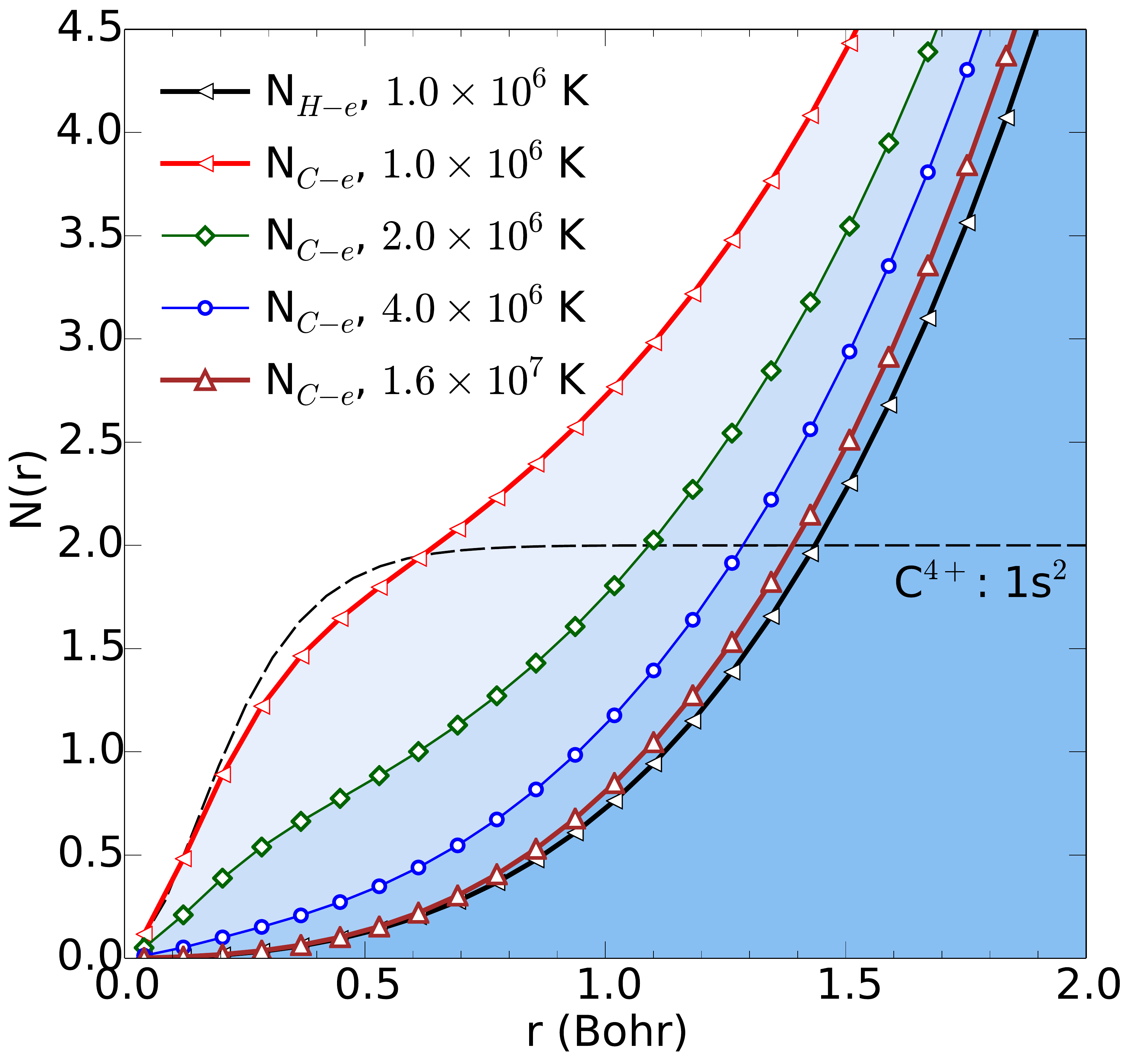}
\caption{\label{nrch} The average number of electrons 
 $N(r)$ around each carbon or hydrogen nucleus 
in CH at 3.15 g/cm$^3$ and a series of temperatures
obtained from the PIMC simulations in this work. 
$N(r)$ is calculated via $N(r) = \left<\sum_{e,I}\theta(r-\left|\vec{r}_e-\vec{r}_I \right|) \right>/{N_I}$,
where the sum includes all electron-carbon ion or electron-hydrogen ion pairs and $\theta$ represents
the Heaviside function. The corresponding profile of the C$^{4+}$ ionization state, calculated with GAMESS, is shown for comparison.}
\end{figure}

Interestingly, our calculations and several other methods and models (Fig.~\ref{fig_huggdp1})
all have a shoulder along the Hugoniot curve
at 4-fold compression. 
This corresponds to a pressure of 10 TPa and temperature of 
5.7$\times10^5$ K, according to the shock compression analysis
of the first-principles EOS data in this work.
The origin of this shoulder may be traced back to the
start of ionization in the carbon 1s shell.
Increasing amounts of carbon 1s electrons are excited at higher temperatures,
as is shown in the $N(r)$ plot of Fig.~\ref{nrch}.
At $10^6$ K, a noticeable amount of excitation can be seen.
The ionization fraction of the carbon 1s shell
grows considerably as temperature
increases even further.
Above 8$\times10^6$ K, 1s ionization is nearly complete and the system
approaches an ideal plasma. Therefore the Hugoniot curves from different
methods and models merge together and reach the ideal Fermi gas limit,
which is consistent with the EOS comparisons and discussions in Sec.~\ref{seceos}.

\begin{figure}
\centering\includegraphics[width=0.5\textwidth]{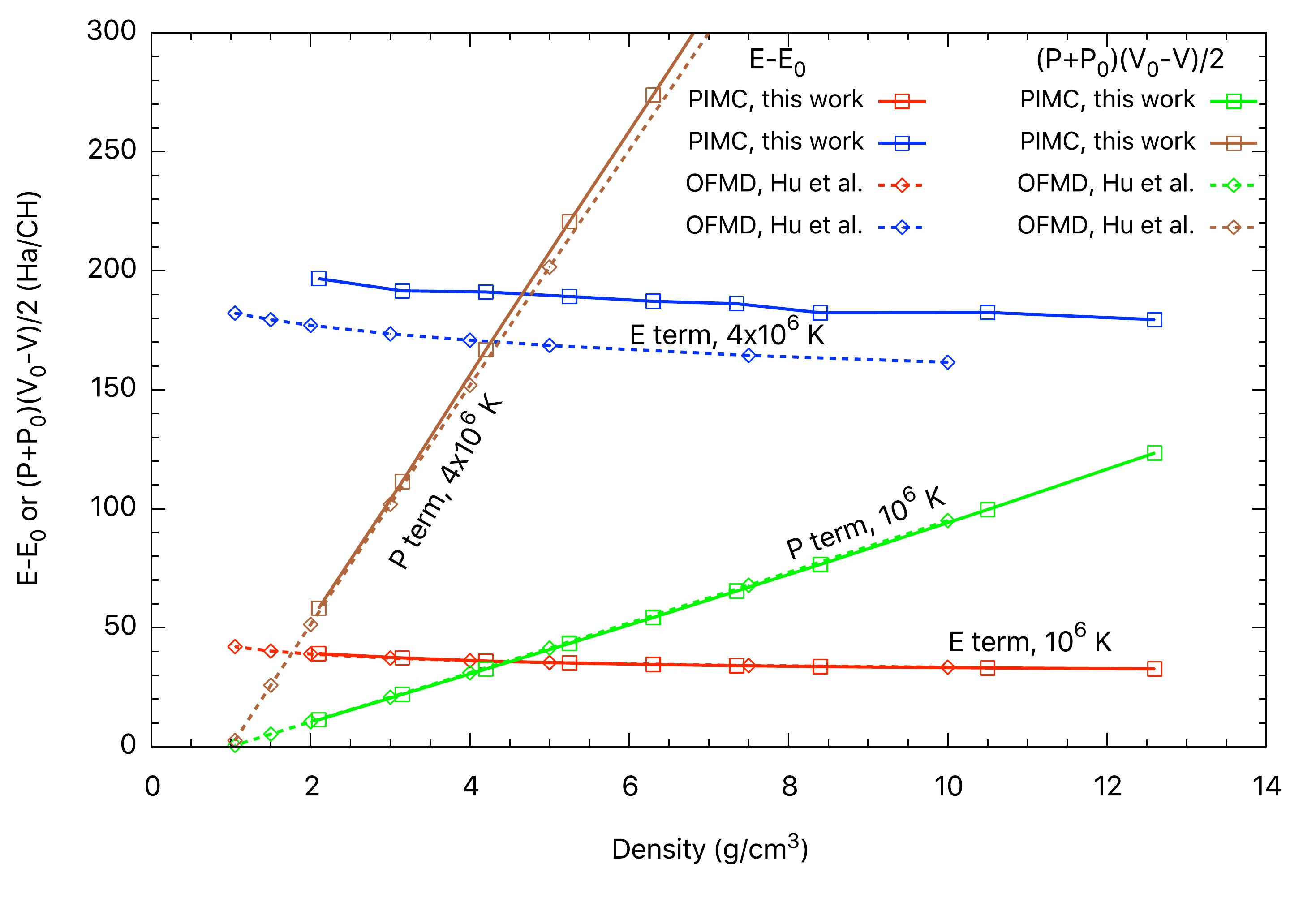}
\caption{\label{pimcvsof} Comparison of the energy and pressure components of the Hugoniot function 
of CH from this work (solid lines and squares) and from the OF-DFT model~\cite{Hu2015}, dashed lines and diamonds) at two different temperatures. 
The initial state correspond to the ambient density of 1.05 g/cm$^3$. 
The densities at which the
energy and pressure terms agree solve Eq.~\ref{eqhug} and may be
realized in experiments.}
\end{figure}

In order to better understand the differences between the Hugoniot curves from
our simulations and the OF-DFT and DFT-MD study in Ref.~\onlinecite{Hu2015},
we compare the two components of the Hugoniot function in Fig.~\ref{pimcvsof}, 
i.e., the internal energy term $E-E_0$ and the pressure term $(P+P_0)(V-V_0)/2$, 
at two different temperatures. 
At $10^6$ K, data in this work and those of Ref.~\onlinecite{Hu2015}
both rely on DFT-MD simulations, thus the energy and pressure values are 
similar. At $4\times10^6$ K, the
OF-DFT pressure is slightly lower than that given by PIMC and the difference between them 
grows larger for higher densities, whereas the internal energy from OF-DFT is  
significantly lower than that of PIMC, resulting in lower densities along the Hugoniot curve.
The differences between the Hugoniot curves from the two methods are similar to those found for Si~\cite{PhysRevB.94.094109}. These differences 
originate from the different treatments of electronic shell ionization effects in 
the two methods---PIMC is a many-body approach that accurately includes shell
effects, while the OF-DFT approach makes use of what is essentially a Thomas-Fermi density functional which is not able to describe the shell structure accurately. 
Therefore, OF-DFT tends to smooth out ionization features near the compression maximum,
leading to a single broad peak instead of the peak-shoulder (for CH) or 
double-peak (for Si) structures predicted by PIMC.

Note that zero-point motion has not been considered in 
our calculation of the reference energy $E_0$ nor in our DFT-MD simulations. 
To estimate its effect, we approximate~\cite{Wallace2011} the 
harmonic zero-point energy with $\delta E=9k_B\Theta_D(V)/8$, 
where $k_B$ is the Bolzmann constant and $\Theta_D(V)$ is 
the volume-dependent Debye temperature, and the corresponding 
pressure $\delta P=9\gamma k_B\Theta_D(V)/8V$, where $\gamma$ is 
the Gr\"{u}eneissen parameter, and add them to our DFT-MD EOS table 
of CH. $\Theta_D(V_0)$ and $\gamma$ are approximated 
by 140 K and 1.2, respectively, by referring to the values of 
plastics in literature~\cite{Amzel1971,Arp1984}. With this zero-point correction, the compression ratio 
along the CH Hugoniot curve is reduced by a small amount of 0.01 for temperatures less than $10^5$ K. 
The maximum correction to our EOS occurs for the highest density of 12$\rho_0$, 
at which $\Theta_D(V)$ reaches a high value of 2761.5 K. However, the zero-point motion
has a negligible effect ($<$0.3\%) when comparing the EOS from 
PIMC and DFT-MD at 10$^6$ K.

\section{Discussion}\label{discuss}
\subsection{Structure of the hydrocarbon fluid}\label{struccharac}

\begin{figure}
\centering\includegraphics[width=0.45\textwidth]{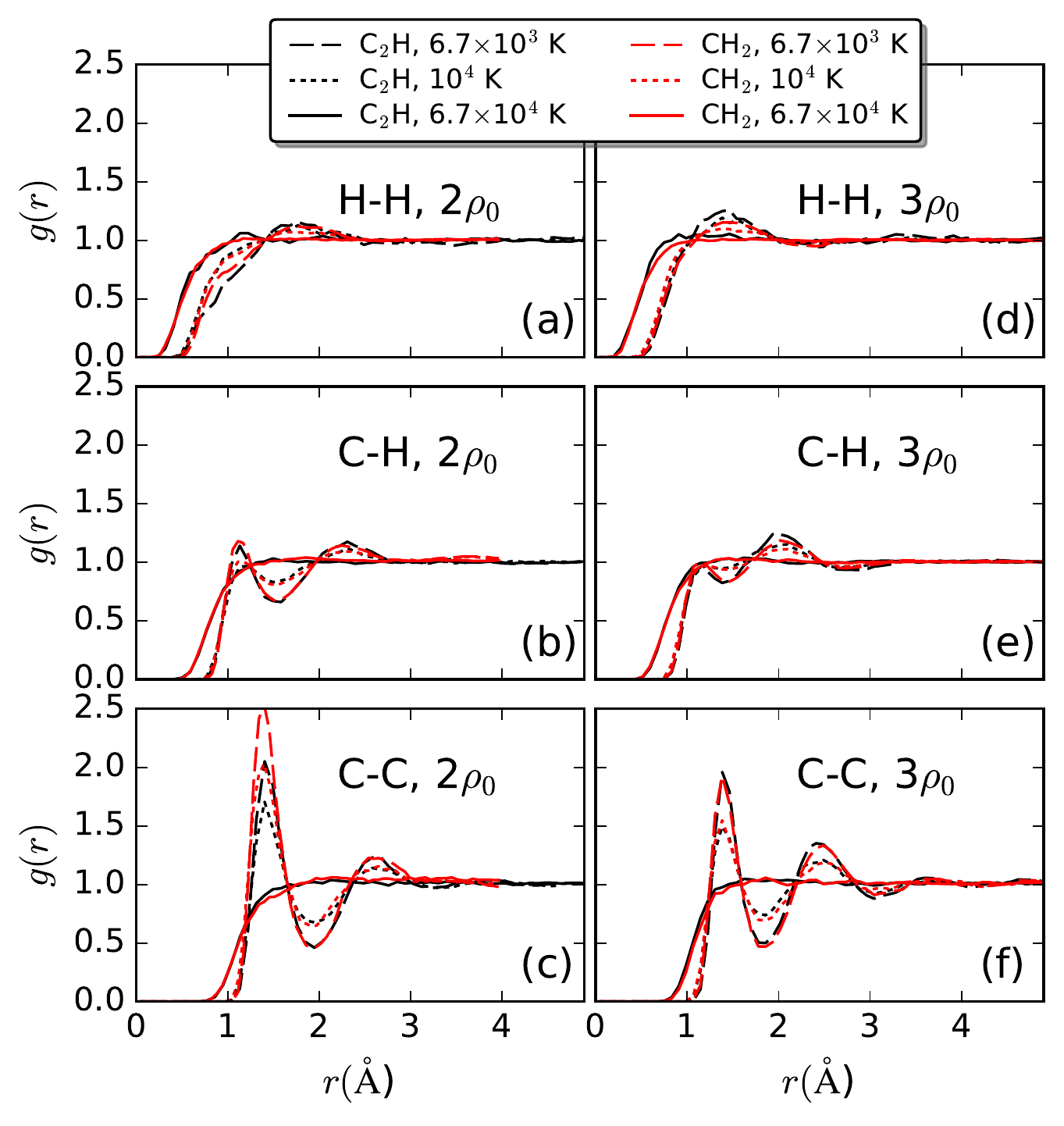}
\caption{\label{fig_gr} Comparison of the nuclear pair correlation
  function obtained from DFT-MD for C$_{2}$H (dark) and CH$_{2}$ (red)
  at two different densities for every material and three temperatures.
The reference density $\rho_0$ is 1.76 and 2.24 g/cm$^3$ for C$_{2}$H and CH$_{2}$,
respectively.}
\end{figure}

In Ref.~\onlinecite{Zhang2017b}, we have shown that the isothermal
isobaric linear mixing approximation is very reasonable for 
hydrocarbons under stellar-core
conditions. The validity of this approximation can be understood from 
analysis of the nuclear pair correlation functions,
as is shown with the high-temperature (6.7$\times10^4$ K) $g(r)$ profiles
in Fig. S2 of Ref.~\onlinecite{Zhang2017b} for CH
and Fig.~\ref{fig_gr} for C$_2$H and CH$_2$.
At these extreme conditions, no C-H bonds exist 
(lifetime shorter than 4.4 fs, see Table~\ref{table:bondlife}), and
the non-existence of any peak structure in the pair-correlation function indicates 
that the system behaves similarly to an ideal yet partially ionized plasma.
This explains the efficacy of the ideal linear mixing assumption manifested in the similarity between the Hugoniot curve as calculated from this approximation 
(using PIMC and DFT-MD EOS for the pure elements~\cite{PhysRevB.84.224109,Benedict2014C}), 
and the Hugoniot curve as calculated 
from our direct first-principles simulations of the mixtures in question~\cite{Zhang2017b}.

\begin{figure}
\centering\includegraphics[width=0.45\textwidth]{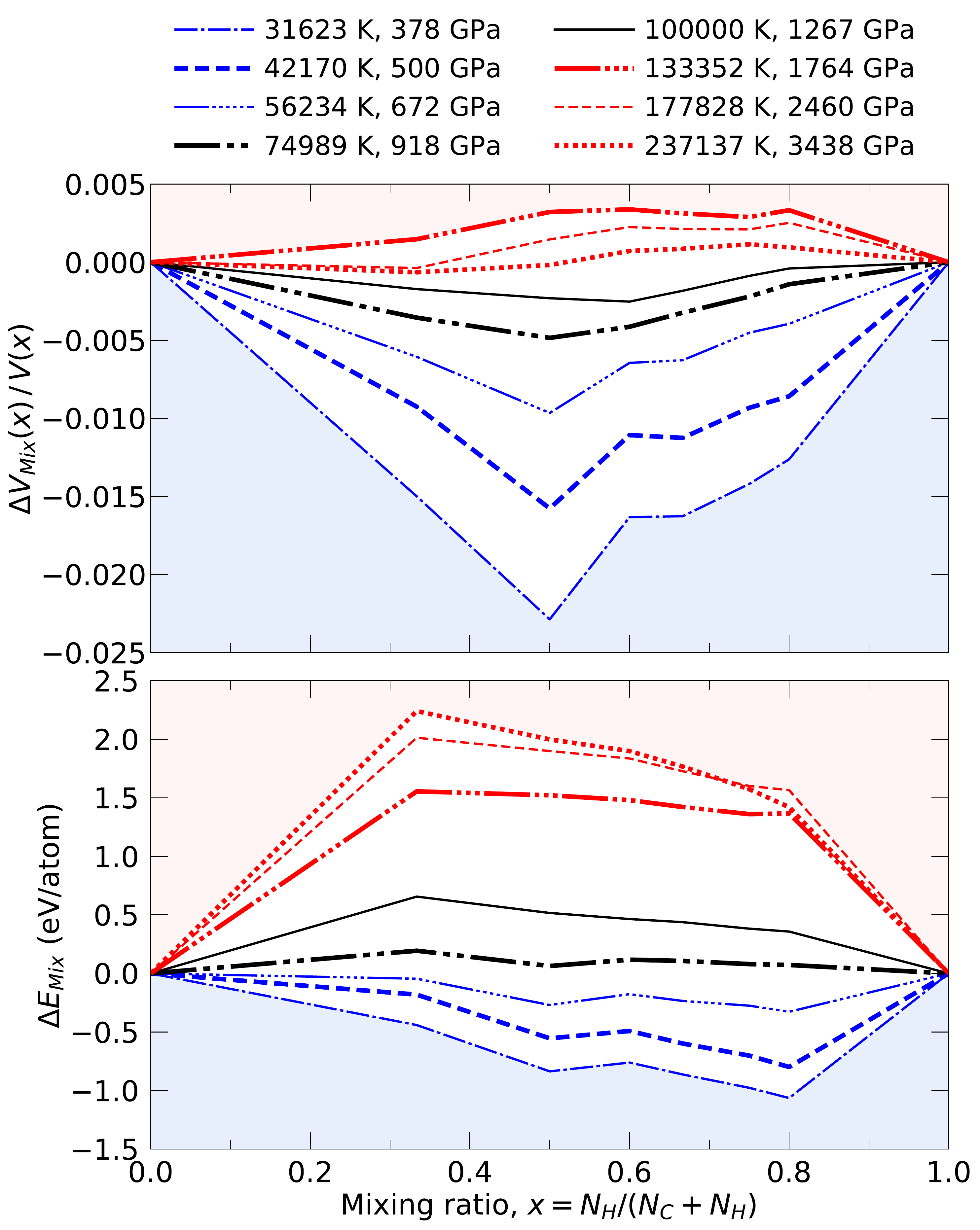}
\caption{\label{dEmixdVmix} The difference in energy (top) and volume (bottom) between
bi-variable spline fitting of our first-principles EOS data of CH and the linear mixing approximation
at a series of $T, P$ conditions along the Hugoniot curve. The symbols are defined as
$\Delta E_{Mix}(x)=E(x)-(1-x)E_\text{C}-xE_\text{H}$ and $\Delta V_{Mix}(x)=V(x)-(1-x)V_\text{C}-xV_\text{H}$.}
\end{figure}

Figure~\ref{dEmixdVmix} compares the EOS of CH from interpolation of 
our first-principles data and that determined by the 
linear mixing approximation at a series of $T, P$ conditions along the Hugoniot curve (Fig.~\ref{fig_huggdp1}). 
In comparison to interpolation of our first-principles EOS data,
linear mixing of pure C and pure H overestimates the volume 
of CH by 2.3\% at
3.2$\times10^4$ K (378 GPa). The volume difference $\Delta V_{mix}$ decreases
to within $\pm$0.5\% at $T>5.6\times10^4$ K, 
which is consistent with the threshold temperature above which 
we see disappearance of peaks in the nuclear-pair correlation function $g(r)$ plots (Fig.~\ref{fig_gr}).
On the other hand, the energy of the linear mixture is higher than
the first-principles value by 1.1 eV/atom at 3.2$\times10^4$~K.
The value of $\Delta E_{mix}$ decreases to 0-0.2 eV/atom at
7.5$\times10^4$ K, and remains less than 2.5~eV/atom
at higher temperatures. The fact that the energy of the linear mixture 
is smaller at the highest temperatures, while pressure is similar, explains 
why linear mixing predicts CH to be stiffer at the compression maximum 
than our direct first-principles 
predictions (Fig.~3 in Ref.~\onlinecite{Zhang2017b}).

At lower pressure and temperature, clear signatures of chemical bonds exist.
Figure~\ref{fig_gr} compares the $g(r)$ profile of C$_2$H and CH$_2$
at two different densities (2$\times$$\rho_0$ and 3$\times$$\rho_0$, with $\rho_0$= 
1.76 and 2.24 g/cm$^3$ for C$_{2}$H and CH$_{2}$, respectively) 
and three different temperatures.
For C-C $g(r)$ functions in Fig.~\ref{fig_gr}(c) and (f), the results show clear peaks and structure 
at both temperatures of 6.7$\times10^3$ and $10^4$ K. These indicate the formation of 
carbon clusters. 
C-H bonds  (Fig.~\ref{fig_gr}(b) and (e)) also exist at these temperatures,
characterized by the peak in $g(r)$ at $r$$\approx$1.15 \AA. 
The C-H
bonds are not very stable and are significantly weakened by thermal and compressional effects,
as the differences in peak height show.
We do not see any evidence of stable H-H bonds even at the the lowest temperature 
($6.7\times10^3$ K) that we considered (Fig.~\ref{fig_gr}(a) and (d)), 
which is consistent with the analysis of other high pressure hydrogen-rich materials \cite{Vorberger2007,Soubiran2015}.
Interestingly, Ref.~\onlinecite{Sun2014} reported the formation of
H$_2$ molecules during the dissociation of silane at low 
densities and temperatures of 1000-4000 K. 
Considerable amount of H$_2$ molecules were also found  
in H$_2$-H$_2$O mixtures at pressures below 35 GPa and 
a temperature of 2000 K~\cite{Soubiran2015}.
Those conditions are much
colder than we have considered in this study. In our simulations,
the lifetime of H-H bonds is generally very short ($<$ 5 fs, see the 
following discussion and Table~\ref{table:bondlife}), which
indicates no stable H$_2$ molecule can be formed.
The peaks in $g(r)$ do not seem to be strongly dependent on the overall C:H ratio
in the system and are consistent with findings in recent DFT-MD simulations~\cite{Sherman2012} of
CH$_4$.

\begin{figure}
\centering\includegraphics[width=0.5\textwidth]{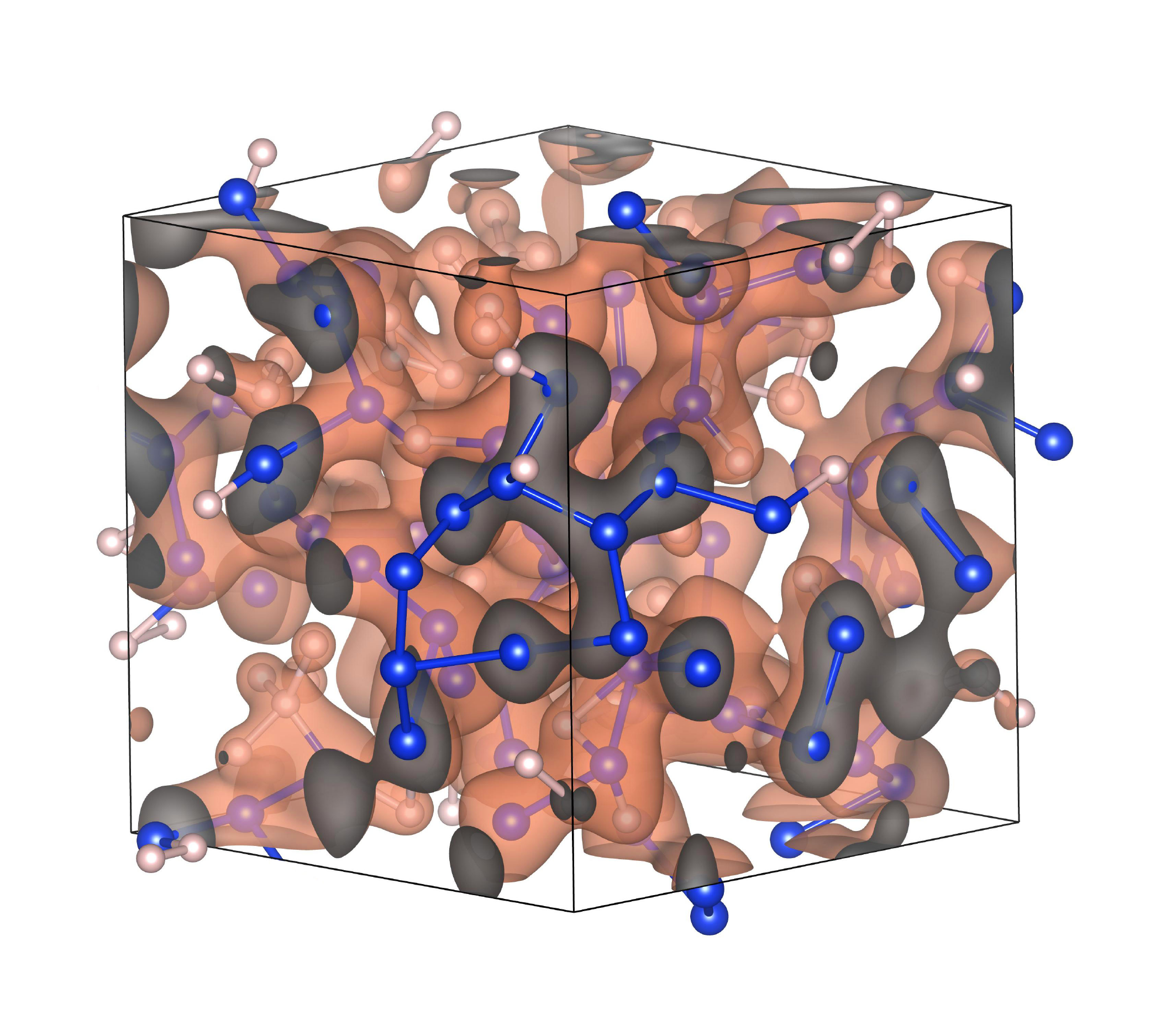}
\caption{\label{chargecar} A snapshot of the electron density profile of CH
in the polymeric state~\cite{Sherman2012}
around thermodynamic equilibrium at 3.15~g/cm$^3$ and 2$\times10^4$ K. 
The C and H atoms are represented
by blue and pink spheres, respectively. 
Electron density which mitigates the bonding effects 
is illustrated by the light brown isosurface.}
\end{figure}

A snapshot from the DFT-MD simulations of CH and its
 electronic density distribution at 3.15 g/cm$^3$ and $2\times10^4$ K is shown in Fig.~\ref{chargecar}.
The disorder in the atomic positions is indicative of the plasma behavior,
wherein ions participate in short-lived chemical bonding.
Detailed structural analysis of the atomic bonding and
lifetime~\footnote{We chose cut-off distances of 1.90 ${\rm \AA}$, 
1.55 ${\rm \AA}$, and 1.00 ${\rm \AA}$ for C-C, C-H, and H-H, 
respectively. These define a pattern of bonds for every configuration 
along the MD trajectory. Then we analyze how long particular double 
and triple bonds exist to determine the lifetime.}
 at this condition (see Table~\ref{table:bondlife}) indicates that C-C clusters and chains 
exist stably for the time scale of 10-100 fs, 
approximately one order of magnitude longer than H-H bonds.
The lifetime of C-H bonds is slightly longer than 10 fs
at $T<10^4$ K and $\rho<3.15$ g/cm$^3$, and is shorter than 10 fs and about twice
the value of H-H bonds at $T>10^4$ K.

\begin{table}
\caption{Comparison of chemical bond lifetime $\tau$ in fs.}
\label{table:bondlife}
\begin{tabular} {ccccccc}
\hline
\hline
$\rho$ (g/cm$^3$) & $T$ (K) & $P$ (Mbar)  & $\tau_\text{C-C}$ & $\tau_\text{C-H}$ & $\tau_\text{H-H}$ & $\tau_\text{C-C-C}$ \\
\hline
2.10   &  6.7$\times10^3$    & 0.44  &  89.62 & 16.68 &  4.60 & 44.12  \\  
2.10   &  1.0$\times10^4$    & 0.61  &  48.77 & 11.36  & 4.00 & 23.92  \\  
3.15   &  6.7$\times10^3$    & 1.76  &  69.72 & 12.63  & 4.18 & 35.67  \\  
3.15   &  1.0$\times10^4$    & 2.04  &  43.13 & 10.19  & 3.80 & 21.72  \\  
3.15   &    2.0$\times10^4$    & 2.89  &  25.63 & 7.58  &  3.15 &  12.93   \\  
3.15   &    5.0$\times10^4$    & 5.55  & 14.63 &  5.06 & 2.28  &  7.55   \\  
3.15   &    6.7$\times10^4$    & 7.19  & 12.44 &  4.34 & 2.06  &  6.58   \\  
3.15   &    1.3$\times10^5$    & 13.4  &  9.50 & 3.30  &  1.60 &   4.88   \\  
3.15   &    2.0$\times10^5$    & 22.1  &  7.47 & 2.68  &  1.32 &   3.85   \\ 
3.15   &    5.1$\times10^5$    & 60.9  &  4.70 & 1.66  &  0.88 &   2.47   \\ 
\hline
\hline
\end{tabular}
\end{table}



Changes in chemical bonding in hydrocarbons have been 
proposed to interpret experimental results along
the Hugoniot curves. For example, Barrios {\it et al}.~\cite{Barrios2010} 
tentatively attributed a slight softening that is observed for polystyrene 
at 2-4 Mbar to the decomposition of chemical bonds. 
Bond dissociation is also included in the SESAME
EOS model for CH at 2-4 Mbar~\cite{sesame}.
Our findings of dramatic decrease in the lifetimes of
C-H bonds and changes in $g(r)$ at 0.4-4 Mbar
are consistent with previous calculations~\cite{Wang2011,Hamel2012,Huser2015,Hu2015}.
However, our analysis shows that the decrease in bond lifetime
is gradual, spaning a few Mbar and tens of thousands of Kelvin.
This indicates that the dissociation of chemical bonds 
is continuous along the Hugoniot curve, instead of
suddenly complete at certain $T$ and $P$.

\begin{figure}
\centering\includegraphics[width=0.5\textwidth]{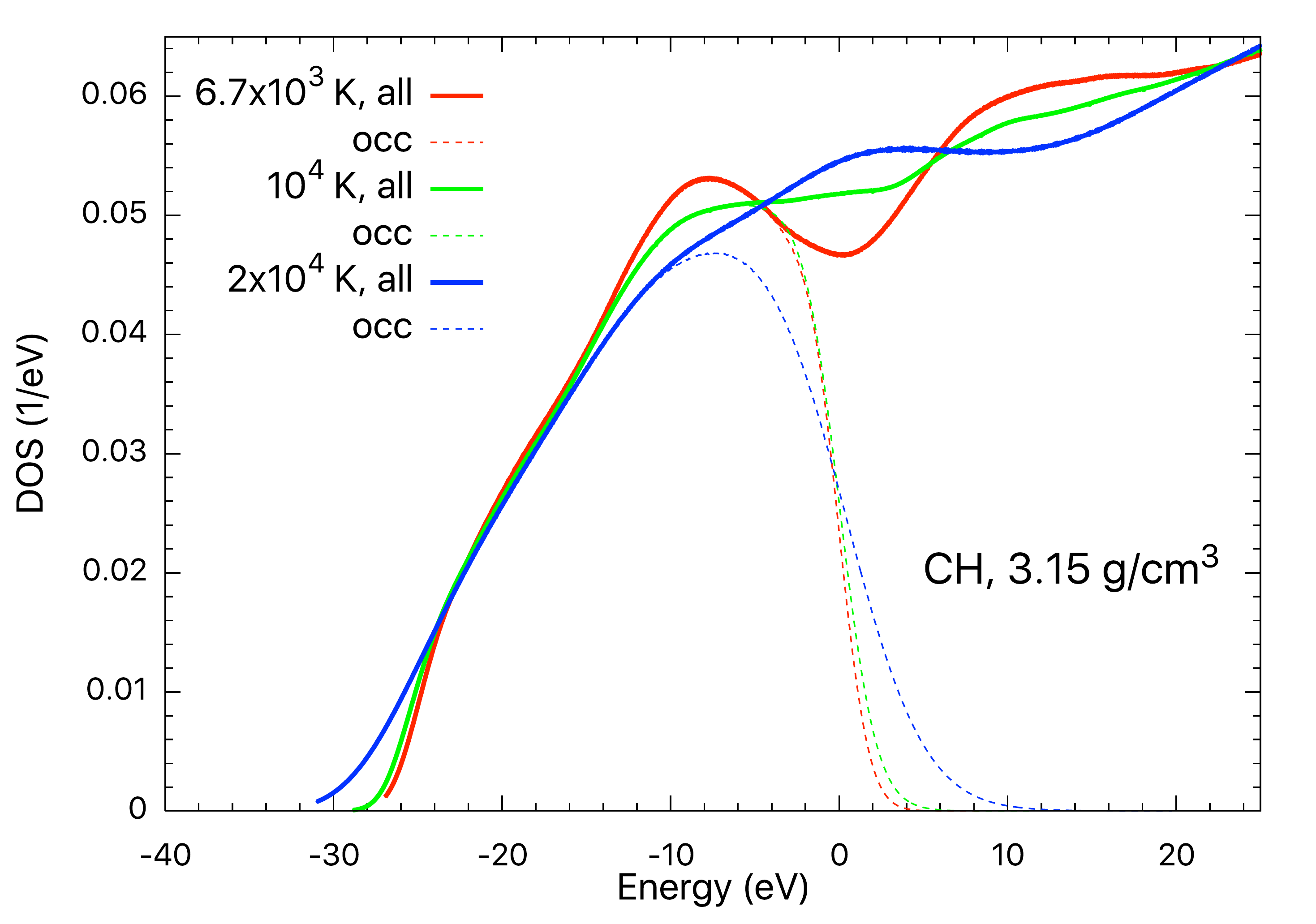}
\caption{\label{fig_dos} Electronic density of states (DOS) of CH at a series of
temperatures and 3.15 g/cm$^3$ based on DFT simulations.
The dashed curves denote the occupied states.
The Fermi energies are aligned at E=0 eV.
A pseudo-bandgap exists at 6.7$\times10^3$ K. With increasing temperature,
the gap is gradually filled. We have compared the results by using the $\Gamma$ point,
and 2$\times$2$\times$2 and 4$\times$4$\times$4 Monkhorst-Pack
$k$ meshes~\cite{Monkhorst1977}, and found very similar relationships between the three DOS curves.}
\end{figure}

We also investigated the electronic density of states of CH
at the same ($\rho, T$) conditions as in the $g(r)$ analyses (Fig.~\ref{fig_gr}).
We find a pseudo-bandgap exists at the valence band maximum at 6.7$\times10^3$ K.
This gap is partially filled, resulting in a more continuous transition between
the valence and the conduction bands, at $10^4$ K, and is completely filled at 2$\times10^4$ K
(see Fig.~\ref{fig_dos}). 
Closure of the gap indicates metalization of the system,
which increases the electrical
conductivity and reflectivity that can be observed in experiments.
The corresponding pressure range (1.5-2.5 Mbar) is in accord with
experimental findings~\cite{Barrios2010,Koenig2003} of
optical reflectivity changes of CH samples.
Note that band gap closure does not necessarily accompany complete chemical decomposition,
because the changes in the lifetimes of the chemical bonds are gradual.
It is therefore not appropriate to equate the origin of reflectivity change
with chemical bond dissociation.
We also do not see the changes in bonding to have an obvious effect on the shape
of the Hugoniot curve.

\subsection{Composition dependence of the Hugoniot curve of GDP}
ICF experiments routinely use
GDP as an ablator material; GDP is mostly hydrocarbon (CH$_{1.36}$)
doped with small amounts of heavier elements, such as O or Ge.
As has been shown in Sec.~\ref{struccharac} and in Ref.~\onlinecite{Zhang2017b},
the linear mixing approximation is a reasonable
way of estimating the EOS and shock compression 
of hydrocarbons. In this section, we apply
this approximation to study the 
shock compression of GDP, compare
with other EOS models, and investigate the
effects of varying hydrogen and oxygen concentrations on the
shock Hugoniot curve.

\begin{figure}
\centering\includegraphics[width=0.45\textwidth]{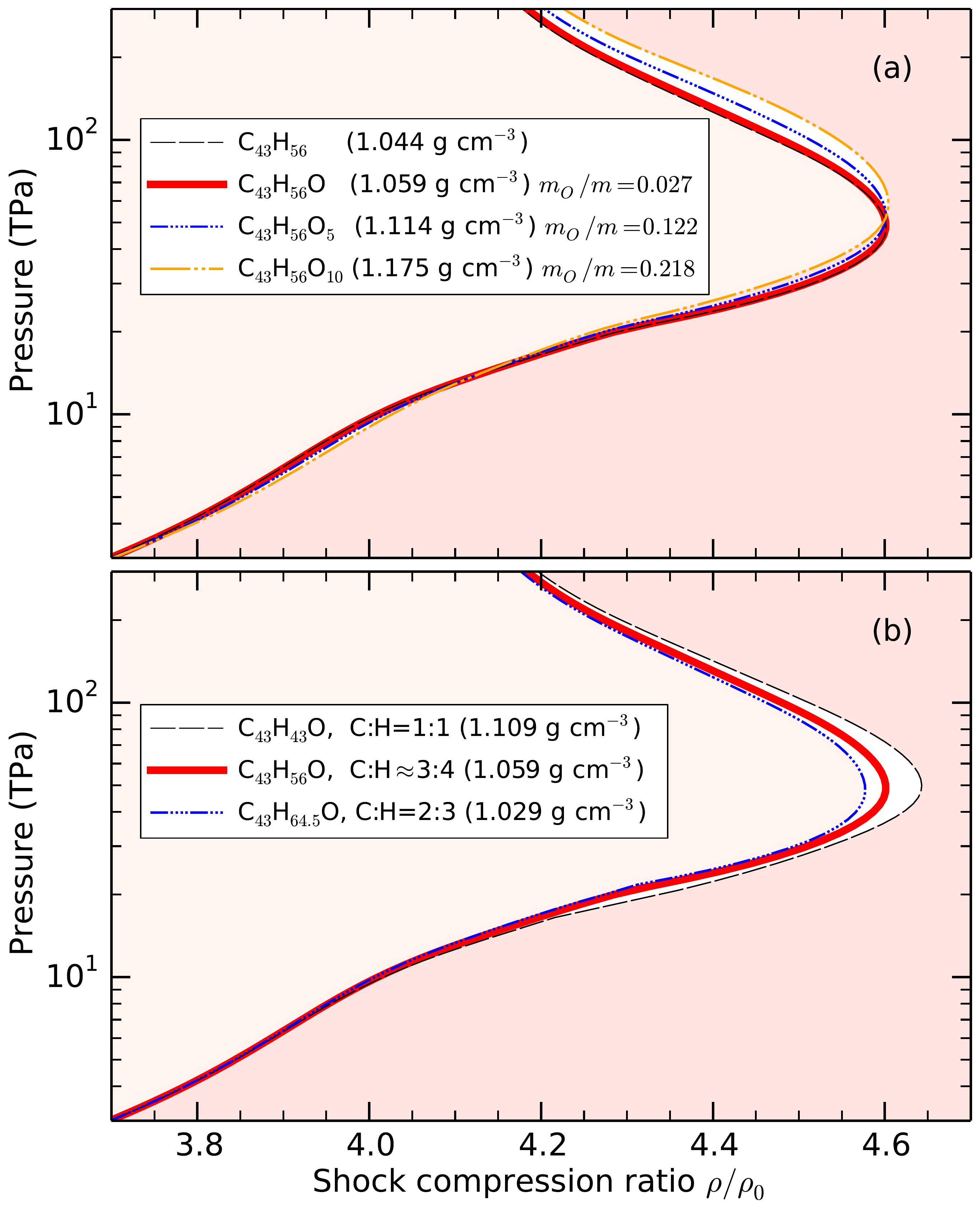}
\caption{\label{fig_huggdp2}Comparison of the effects of varying oxygen (a) and hydrogen (b) content on the shock Hugoniot curve of GDP.}
\end{figure}

Figure~\ref{fig_huggdp1} shows the Hugoniot curve of GDP (C$_{43}$H$_{56}$O), 
in comparison with those given by
other models
and that of CH. The initial density of GDP is set to 1.05 g/cm$^3$, 
as relevant to polystyrene and to the hydrocarbon materials used in recent laser shock
experiments~\cite{Barrios2012}. 
The initial energy $E_0$ is determined
by approximating GDP as an ideal mixture of 
three polymers---PAMS (C$_9$H$_{10}$), polyethylene (C$_2$H$_4$) and 
polyvinylalcohol (C$_2$H$_4$O)---which 
allows for a higher flexibility in the composition of GDP.
The energy of each polymer is determined as described in 
the supplementary material of Ref.~\onlinecite{Zhang2017b}.
We determine the energy of diamond, 
isolated H$_2$ and O$_2$ molecules using 
DFT calculations and combine them with 
tabulated thermochemical data 
on the enthalpy of combustion~\cite{Roberts1951,Walters2000}
to estimate the energy of these polymers.
As a proof of concept, we compare the energy of coronene C$_{24}$H$_{12}$ 
determined as a combination of PAMS and polyethylene to the thermochemical data
 and find very small difference (less than 15 mHa/carbon). 
Initial densities in Fig.~\ref{fig_huggdp2} are determined 
in the ideal mixture approximation, using the density of PAMS (1.075 g/cm$^3$), 
polyethylene (0.95 g/cm$^3$) and polyvinylalcohol (1.19 g/cm$^3$).

Figure~\ref{fig_huggdp1} shows the Hugoniot curve of GDP obtained
from our first-principles EOS. It is slightly stiffer than that of CH at low
pressures ($P<5$ Mbar) and near the compression maximum. 
Similar trends are found in the results of the QSEM model~\cite{Colin2016}.
For compression ratios between 3.3-4.3,
a shoulder develops along the Hugoniot curve.
The shoulder structure is also found  along the Hugoniot curve of LEOS 5400 GDP,
which exhibits higher compressibility than that predicted by the first-principles calculations in this work.  This may be traced to the start of ionization of the 1s electron shell of carbon, 
which leads to the shoulder,
and the much lower Gr\"{u}eneisen $\gamma$ used in LEOS 5400 than in LEOS 5112 for compression ratios larger than 3.5,
which causes the softer behavior predicted by LEOS 5400.

Considering that the chemical composition of GDP ablators varies~\cite{Barrios2012,Colin2016,Hamel2012,Sterne2016}, it is thus useful to compare the Hugoniot curves of GDP with different
C:H ratios and oxygen contents. We consider three 
oxygen mass percentages of 2.7\%, 12.2\%, and 21.8\%,
and C:H ratios of 1:1, 1:1.33, and 1:1.5. 
A comparison of the Hugoniot curves
is shown in Fig.~\ref{fig_huggdp2}. 
With the addition of oxygen,
the pressure at the maximum compression increases, while the 
compression ratio does not show any observable change. 
The effect of oxygen can be understood
from the fact that its 1s electrons are more strongly bound to the
oxygen nuclei,
which requires a higher temperature for ionization. 
Changes to the maximum compression ratio
are insignificant within the range of oxygen content that we consider, 
because the carbon 1s electrons dominate over those
of oxygen.
Increasing the amount of hydrogen leads to a decrease
by 0.1 in the compression maximum when the C:H ratio decreases from 1:1 to 1:1.5,
as shown in Fig.~\ref{fig_huggdp2}(b). This is the same trend with composition
that has been seen in Fig.~3 of Ref.~\onlinecite{Zhang2017b}. The pressure at the compression
maximum is not affected.

\section{Conclusions}\label{conclusion}
In this work, we presented the results of PIMC and DFT-MD simulations 
of a series of hydrocarbons. We obtained accurate internal energies and pressures 
from temperatures of 6.7$\times10^3$ K to 1.3$\times10^8$ K. 
PIMC and DFT-MD were shown to be consistent at $10^6$ K,
typically to within 1 Ha/carbon in energy and 3\% in pressure.
This cross-validates the use of both methods at the temperature
of $10^6$ K. We used these results to evaluate some of the detailed assumptions made in a recent EOS model for CH, LEOS-5112.

We investigated the principal Hugoniot curves using the obtained
EOS data and found a maximum compression of 4.7, which is similar to that
predicted by LEOS 5112 and 5400 but larger than SESAME 7593 and OF-DFT
predictions. We expect future experiments will test this prediction.

We demonstrated the validity of the linear mixing approximation
in obtaining the EOS and shock Hugoniot curves of hydrocarbons.
This can be explained by 
the unstable, short-lived 
C-H chemical bonds (lifetime $\tau<4.4$ fs) for 
temperatures greater than $6.7\times10^4$ K. 
The nuclear-pair correlation function $g(r)$ of hydrocarbons resembles that of a simple atomic liquid 
at the higher temperatures we considered. 
At lower temperatures, $g(r)$ as well as
bond lifetime analysis of hydrocarbon systems 
show the possible existence of stable C-C bonds
with lifetimes of 10-100 fs, 
weak C-H bonds with lifetimes of 4-16 fs, 
and no signature of stable H-H bonds.

By applying the linear mixing approximation, we investigated
the Hugoniot curves of GDP as a function of oxygen content
and C:H ratios. We found that the compression maximum remains unchanged
when varying the oxygen mass percentage between 0 and 21.8\% while
the pressure increases by about 0.1 Gbar. When the C:H ratio
decreases from 1:1 to 1:1.5, the shock compression maximum decreases by 0.1 while
the pressure, which is determined by the 1s ionization of carbon,
does not change.

Our results provide a benchmark for future theoretical investigations of the EOS 
of hydrocarbons, and should be useful for informing on-going dynamic compression experiments aimed at reaching Gbar conditions.

\section{Supplementary Material}
See the supplementary material for the tables of EOS data of C$_2$H, CH,
 C$_2$H$_3$, CH$_2$,
 CH$_3$, and CH$_4$ considered in this study.

\begin{acknowledgments}
  
  This research is supported by the U. S. Department of Energy, grant
  DE-SC0010517, DE-SC0016248, and DE-NA0001859. 
Computational support was provided by the
  Blue Waters sustained-petascale computing project 
  (NSF ACI 1640776), which 
  is supported by the National Science Foundation 
  (awards OCI-0725070 and ACI-1238993) and the state of Illinois. 
  Blue Waters is a joint effort of the University of Illinois at 
  Urbana-Champaign and its National Center for Supercomputing Applications. 
S.Z. is partially supported by the PLS-Postdoctoral Grant of the 
Lawrence Livermore National Laboratory.
  S.Z. and B.M. thank Benjamin D. Hammel for putting forward the interesting question about 
bond dissociation. S.Z. appreciate Miguel Morales for helpful discussions.
This work was in part performed under the auspices of the U.S. Department of Energy by Lawrence Livermore National Laboratory under Contract No. DE-AC52-07NA27344.
  
\end{acknowledgments}


%

\end{document}